\documentclass[11pt,a4paper]{article}
\usepackage{amsmath}
\usepackage{amsfonts}
\usepackage{color}
\usepackage{graphicx}
\usepackage{dcolumn}
\usepackage{bm}
\usepackage[
bookmarks,bookmarksnumbered,colorlinks=true,anchorcolor=blue,
linkcolor=blue,urlcolor=blue,citecolor=blue,
breaklinks=true]{hyperref}
\usepackage{geometry}
\usepackage{hep}
\usepackage{extarrows}
\usepackage{authblk}
\usepackage{amssymb}
\usepackage[utf8]{inputenc}
\usepackage[numbers,sort&compress]{natbib}
\usepackage{cite}
\usepackage{pifont}
\usepackage{tikz}
\usetikzlibrary{shadings,intersections,snakes}
\usepackage{verbatim}
\numberwithin{equation}{section}   

\date{\today}

\begin{document}


\title{\bf Schwinger-Keldysh effective action for relativistic Brownian particle in AdS-CFT}

%
%

\author[]{Yanyan Bu \thanks{yybu@hit.edu.cn}~}

\author[]{Biye Zhang \thanks{zhangbiye@hit.edu.cn (correspondence author)}}


\affil[]{\it School of Physics, Harbin Institute of Technology, Harbin 150001, China}

\maketitle

\setcounter{page}{0}
\thispagestyle{empty}

\begin{abstract}

We compute Schwinger-Keldysh effective action for a relativistic heavy quark (with constant background velocity) in strongly coupled $\mathcal N=4$ supersymmetric Yang-Mills plasma. The holographic dual description involves a noisy trailing string moving in Schwarzschild-AdS$_5$ black brane. The noise is caused by Hawking radiation emitted from string worldsheet horizon. Besides quadractic terms, the effective action contains cubic interactions, which are entirely induced by the constant background velocity. The nonlinear fluctuation-dissipation relations for three-point functions are discussed based on holographic results.

%

\end{abstract}


\newpage

\tableofcontents

\allowdisplaybreaks

\flushbottom

\section{Introduction}

Brownian motion of a heavy particle in a thermal bath has become a textbook example \cite{Calzetta-Hu2009,Kamenev2011} for understanding fundamental aspects of non-equilibrium statistical mechanics. Langevin equation is perhaps the simplest {\it effective} treatment over complicated dynamics of Brownian motion:
\begin{align}
M_{\rm kin}\frac{d^2}{dt^2} q(t) + \eta \frac{d}{dt} q(t)  = \xi(t), \label{Langevin_eq}
\end{align}
where $q(t)$ is position of Brownian particle, $\eta$ represents the drag force, and the stochastic variable $\xi(t)$ stands for random force felt by the Brownian particle. Usually, the random force is assumed to be white noise:
\begin{align}
\langle \xi(t)\rangle=0, \qquad  \langle \xi(t) \xi(t^\prime) \rangle = 2 T \eta \delta(t-t^\prime),
\end{align}
where the strength of noise-noise correlator is related to drag force through fluctuation-dissipation theorem. Here, $T$ is the temperature of thermal bath.

\begin{figure}[htbp!]
	\centering
	\begin{tikzpicture}[]
	\draw[very thick] (-5,-0.2)--(5,-0.2);
	\node[above] at   (0,0.2) {$U(t_f,t_i)$};
	\draw[very thick] ( 4.98,-0.2)--(4.98, 0.2);
	\node[above] at   (5,0.2) {$t_f$};
	\draw[very thick] (-5, 0.2)--(5, 0.2);
	\node[below] at   (0,-0.2) {$U^\dagger(t_f,t_i)$};
	\draw[ ->,very thick] (-1,0.2)--(0.14,0.2);
	\draw[ <-,very thick] (0,-0.2)--(1,-0.2);
	\draw[fill]  (-5,-0.2) circle [radius=0.1];
	\draw[fill]  (-5,0.2) circle [radius=0.1];
	\node[left]  at (-5,0) {$\rho_0$};
	\node[above] at (-5.3,0.2) {$t_i$};
	\end{tikzpicture}
\caption{The SK closed time contour: $\rho_0$ is the initial density matrix, and $U(t_f,t_i)$ is the time-evolution operator from initial time $t_i$ to final time $t_f$.} \label{SK contour}
\end{figure}
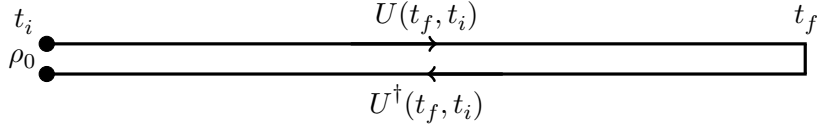

The Langevin theory of Brownian motion \eqref{Langevin_eq} could be derived from real time path integral on closed time contour \cite{Caldeira:1982iu,Feynman:1963fq} (see also \cite{Son:2009vu,Akamatsu:2012vt}). The latter is also called Schwinger-Keldysh (SK) formalism \cite{Keldysh:1964ud,Schwinger:1960qe,Chou:1984es}, which provides a unified framework for describing quantum many-body system in and out of equilibrium \cite{Chou:1984es}. Within the SK formalism, the quantum system {\it effectively} evolves forward (from initial time $t_i$ to final time $t_f$) and then backward (from $t_f$ to $t_i$), forming a closed time contour, as depicted in Figure \ref{SK contour}.
Moreover, the SK formalism systematically incorporates both fluctuations and dissipations, which is achieved by the doubling of degrees of freedom: $q \to (q_1, q_2)$, where the subscripts $1,2$ denote variables on the upper and lower branches of the SK contour in Figure \ref{SK contour}.

Brownian motion involves two key concepts in non-equilibrium statistical mechanics: closed system versus open system \cite{Calzetta-Hu2009}. Thinking of the Brownian particle and the thermal bath as a whole forms a closed system, denoted as $\{q,\Phi\}$. Dynamics of this closed system is presumably described by an action:
\begin{align}
S_C= S_p[q] + S_{th} [\Phi] + S_{int}[q,\Phi], \label{action_closed}
\end{align}
where $S_p[q]$ is action for the Brownian particle, $S_{th}[\Phi]$ describes the microscopic theory of the constitutes (collectively denoted as $\Phi$) for the thermal bath, and $S_{int}[q,\Phi]$ is responsible for the interactions between the Brownian particle and constituents of the thermal bath. The partition function for the closed system is
\begin{align}
Z = \int [Dq] [D\Phi] e^{i S_C}. \label{Z_closed}
\end{align}
However, the microscopic theory \eqref{action_closed} contains too many degrees of freedom and is very hard (if not impossible) to study in general. On the other hand, the main interest is on the behavior of the Brownian particle. This motivates the concept of open system: here, it is the Brownian particle itself, denoted as $\{q\}$. Moreover, this open system will be described by some {\it effective} theory, in which the only degree of freedom is the Brownian particle. In principle, the desired effective theory could be obtained by integrating out, in \eqref{Z_closed}, all the degrees of freedom for the thermal bath. Then, the partition function \eqref{Z_closed} is equivalently expressed as a path integral over position of the Brownian particle only,
\begin{align}
Z = \int [Dq]  e^{i S_{eff}[q]}. \label{Z_open}
\end{align}
Certainly, in order to systematically incorporate fluctuations and dissipations, such a Wilsonian renormalization group (RG) flow should be carried out over the SK closed time contour. The resultant SK effective action $S_{eff}[q]$ will encode information of the thermal bath, interactions between the Brownian particle and the thermal bath, as well as the state.

In practice, the ``integrating out'' procedure outlined above is very challenging to carry out\footnote{In simple examples, the degrees of freedom for the thermal bath could be integrated out, giving linear Langevin equation \cite{Caldeira:1982iu} and its nonlinear extensions \cite{Chakrabarty:2018dov,Chakrabarty:2019qcp}. In these examples, microscopic details of the thermal bath are simplified as a collection of harmonic oscillators. Moreover, the interaction between the Brownian particle and bath oscillator was assumed to be either Gaussian \cite{Caldeira:1982iu} or weakly cubic \cite{Chakrabarty:2018dov,Chakrabarty:2019qcp}.}. It is thus natural to turn to effective field theory (EFT) approach, and construct the effective action based on symmetry principle. In recent years, in order to tackle with dissipation, the SK formalism has been applied to formulate an EFT for dissipative hydrodynamics \cite{Endlich:2012vt,Kovtun:2014hpa,Nicolis:2013lma,Harder:2015nxa,Grozdanov:2013dba,
Crossley:2015evo,Glorioso:2017fpd,Haehl:2015foa, Haehl:2015uoc,Haehl:2018lcu,
Jensen:2017kzi,Chen-Lin:2018kfl,Baggioli:2020haa,Jain:2020hcu} (see \cite{Glorioso:2018wxw} for a pedagogical review). This significant achievement for hydrodynamic EFT has been largely inspired by the AdS-CFT correspondence \cite{Maldacena:1997re,Gubser:1998bc,Witten:1998qj}, which provides a tractable framework for studying dynamics of strongly coupled large $N_c$ (the number of colors) gauge theory via weakly coupled gravitational theory in asymptotic AdS space.

Brownian motion of heavy quark became an important example in revealing structure of thermal noise in AdS-CFT. A heavy quark in AdS-CFT \cite{Herzog:2006gh,CasalderreySolana:2006rq,Gubser:2006bz,Casalderrey-Solana:2007ahi,
Gubser:2006nz} is represented as an open string stretching from the horizon up to a probe brane, and the string's endpoint on the probe brane is quark's position\footnote{Interestingly, the quark string model was recently taken to understand chaos and realize holographic dual of SYK model \cite{Murata:2017rbp,deBoer:2017xdk,Cai:2017nwk}.}. The stochastic Langevin equation \eqref{Langevin_eq} was derived in \cite{Son:2009vu} (see also \cite{deBoer:2008gu}) via Krustal extension \cite{Herzog:2002pc} of finite temperature AdS-CFT \cite{Son:2002sd}, in which fluctuation-dissipation theorem turns out to be an outcome of the correspondence, rather than being imposed by hand \cite{Son:2002sd}. This study was further extended in a number of aspects: the heavy quark could be relativistic \cite{Giecold:2009cg,CasalderreySolana:2009rm}, the environment would be in a non-equilibrium state corresponding to time-dependent AdS black hole \cite{CaronHuot:2011dr}, the dissipation (the noise) would be nonlinear (non-Gaussian) \cite{Chakrabarty:2019aeu}, etc. Particularly, the latter adopted the holographic SK contour \cite{Glorioso:2018mmw} (see \cite{Skenderis:2008dh,Skenderis:2008dg,deBoer:2018qqm} for an alternative holographic prescription for SK contour) and derived the effective action for a nonlinear Langevin theory \cite{Chakrabarty:2018dov,Chakrabarty:2019qcp}. Over the past two years, the holographic SK contour \cite{Glorioso:2018mmw} attracted a lot of attention in the study of SK field theory \cite{Chakrabarty:2019aeu,Jana:2020vyx,Chakrabarty:2020ohe,Loganayagam:2020eue,
Loganayagam:2020iol,Ghosh:2020lel,Bu:2020jfo,Bu:2021clf,He:2021jna}.

In this work, we will employ the proposal \cite{Glorioso:2018mmw} and explore leading order nonlinear corrections to relativistic Langevin theory of \cite{Giecold:2009cg,CasalderreySolana:2009rm}. A heavy quark in a steady state motion (i.e., with a constant background velocity) corresponds to a trailing string solution in the target AdS black brane \cite{Herzog:2006gh,CasalderreySolana:2009rm}. The trailing string solution induces an event horizon on the worldsheet, which is different from that of target space. As clarified in \cite{Giecold:2009cg,CasalderreySolana:2009rm}, it is the Hawking radiation behind this worldsheet horizon that causes fluctuations on the heavy quark. Thus, the emergent AdS black brane on the worldsheet will be doubled in the spirit of \cite{Glorioso:2018mmw}. Perturbing around the trailing string background, we expand the string's Nambu-Goto action in amplitude of string perturbation: besides quadratic terms, the action also contains cubic interactions. The cubic terms are entirely induced by quark's background velocity, and are not present in \cite{Chakrabarty:2019aeu}. Intriguingly, these cubic terms explicitly break worldsheet time-reversal ($\mathcal{T}$) symmetry. The cubic interactions generate nonlinear corrections to relativistic Langevin theory of \cite{Giecold:2009cg,CasalderreySolana:2009rm}, which do not satisfy Kubo-Martin-Schwinger (KMS) conditions for three-point functions of medium force \cite{Wang:1998wg,Crossley:2015evo}. This supports the claim that the fluctuations and dissipations caused by the worldsheet horizon shall be of non-thermal nature \cite{Giecold:2009cg}.

%
%
%
%
%
%
%

The rest of this paper will be structured as follows. In section \ref{holo_setup}, we present the relativistic trailing string in AdS$_5$ black brane, which is the holographic dual of a relativistic heavy quark in $\mathcal N=4$ SYM plasma. We also analyze nonlinear dynamics for classical string perturbation around the trailing string background. In section \ref{linearized_string_solution}, we analyze the solution for classical string perturbation on the holographic SK contour associated with the string worldsheet horizon. In section \ref{SK_effetive_action}, we compute the SK effective action for the relativistic heavy quark, up to cubic order in the quark's position. At quadratic order, we calculate the SK effective action first analytically in the hydrodynamic limit and then numerically for arbitrary value of frequency. At cubic order, we compute the SK effective action analytically in small frequency regime. In section \ref{summary}, we make a summary of this work and outlook future directions. Appendices \ref{S_NG_2nd_3rd_Schw}, \ref{non-commute} and \ref{KMS_3point_review} supplement further calculational details.

\section{Holographic setup} \label{holo_setup}

\subsection{General statement}

The thermal bath is holographically described by Schwarzschild-AdS$_5$ black brane (the AdS radius has been set to unity):
\begin{align}
ds^2_{\rm TG}=g_{MN} dX^M dX^N =\frac{1}{z^2} \left[ \frac{dz^2}{f(z)} - f(z) dt^2 + \delta_{ij} dX^i dX^j \right], \qquad i,j=1,2,3, \label{target_AdS5}
\end{align}
where $f(z)= 1-z^4/z_h^4$. The Hawking temperature is $T= 1/(\pi z_h)$, which is also the temperature of thermal bath. According to AdS/CFT correspondence, dynamics of a heavy quark is determined by an open string moving in the target space \eqref{target_AdS5}, see \cite{Herzog:2006gh,Gubser:2006bz,CasalderreySolana:2006rq}. The open string stretches from the event horizon $z=z_h$ up to a probe brane placed near the AdS boundary, with string's endpoint on the probe brane identified with position of the heavy quark. In this work, we will focus on an infinitely heavy quark, which corresponds to placing the probe brane at the AdS boundary $z=0$.

Dynamics of an open string is described by Nambu-Goto action
\begin{align}
S_{\rm NG}= -\frac{1}{2\pi \alpha^\prime} \int d^2\sigma \sqrt{-h(X)}, \label{S_NG}
\end{align}
where $h$ is determinant of the induced metric $h_{ab}$ on string worldsheet:
\begin{align}
ds^2_{\rm WS}= h_{ab}d\sigma^a \sigma^b = g_{MN}\frac{\partial X^M}{\partial \sigma^a} \frac{\partial X^N}{\partial \sigma^b}. \label{hab_general}
\end{align}
In static gauge, the worldsheet coordinate is $\sigma^a\equiv (\sigma, \tau) = (z,t)$, and embedding of the open string in the target space \eqref{target_AdS5} is specified by the spatial coordinates $X^i(\sigma^a)$.

In presence of an external Maxwell field, the Nambu-Goto action \eqref{S_NG} should be supplemented by a boundary term:
\begin{align}
S_{\rm bdy} = \int d\tau A_M(X) \frac{dX^M}{d\tau} \bigg|_{\rm bdy} = - \int d\tau X^M \partial_\tau A_M(X)\big|_{\rm bdy}, \label{S_bdy}
\end{align}
where in the second equality we ignored a total time-derivative term.

%

The partition function for the bulk theory is
\begin{align}
Z_{\rm bulk} = \int [DX] [D g_{MN}] e^{i S_{\rm bulk}[X, \, g_{MN}]},
\end{align}
where $S_{\rm bulk} $ is the total action for the bulk theory. In probe limit, the target space \eqref{target_AdS5} does not fluctuate. Then, the bulk partition function $Z_{\rm bulk}$ gets reduced into that of an open string in the curved background spacetime \eqref{target_AdS5}:
\begin{align}
Z_{\rm bulk} \simeq  Z_{\rm string}= \int [D X] e^{i S[X]}, \label{Z_string}
\end{align}
where $S$ is the total string action
\begin{align}
S= S_{\rm NG} + S_{\rm bdy}.
\end{align}
It will be clear that the string embedding profile $X$ is a functional of quark's position $q$, i.e., $X=X[q]$. Thus, the bulk path integral \eqref{Z_string} will be eventually cast into a path integral over the position $q$. We will work in the saddle point approximation:
\begin{align}
Z_{\rm string}= \int [Dq] e^{i S[X[q]]},
\end{align}
where $S[X[q]]$ is the on-shell classical string action. The AdS-CFT conjectures that $Z$ of \eqref{Z_closed} is equivalent to $Z_{\rm bulk}$. Thus, in the probe limit, the on-shell string action $S[X[q]]$ will be identified with the effective action $S_{eff}[q]$ for Brownian particle in plasma medium. Therefore, derivation of $S_{eff}[q]$ boils down to solving the classical equation of motion (EOM) for an open string in Schwarzschild AdS$_5$.

\subsection{Trailing string in Schwarzschild-AdS$_5$ and holographic SK contour} \label{review_trailing_string}

We are interesting in steady state motion of a heavy quark in $\mathcal N=4$ SYM plasma. On the gravity side, this corresponds to a trailing string in Schwarzschild-AdS$_5$ \cite{Herzog:2006gh,Gubser:2006bz}. Consider a ``steady'' profile for the string embedding in the target space:
\begin{align}
\bar X^1= \beta t + \beta \zeta(z), \qquad  \bar X^2= \bar X^3 = 0, \label{trailing_ansatz}
\end{align}
where we use ``bar'' to denote the steady state configuration. 
%
%
The trailing string solution was presented in \cite{Herzog:2006gh,Gubser:2006bz}:
\begin{align}
\zeta(z) = \frac{1}{2\pi T} \left[ \tan^{-1}(z\pi T) - \tanh^{-1}(z\pi T) \right]. \label{trailing_zeta}
\end{align}
For the configuration \eqref{trailing_zeta}, the energy and momentum will flow down the string into the horizon of target space. Therefore, such a trailing string solution has to be maintained by a constant external electric field \cite{Herzog:2006gh}. The relation between the velocity $\beta$ and electric field $E$ is determined by the demand that the string's action is time independent and real \cite{CasalderreySolana:2009rm}:
\begin{align}
\beta= \frac{2E\alpha^\prime}{\sqrt{\pi^2 T^4 +4 E^2 \alpha^{\prime2}}}
=\frac{2E}{\sqrt{\lambda_{\rm tH}} \pi T^2} \frac{1}{\sqrt{1+\frac{4E^2}{\lambda_{\rm tH} \pi^2 T^4}}}, \label{beta-E}
\end{align}
where $\lambda_{\rm tH} = 1/\alpha^{\prime2}$ is the 't Hooft coupling constant \cite{Maldacena:1997re}.

Plugging the trailing string solution \eqref{trailing_ansatz} and \eqref{trailing_zeta} into \eqref{target_AdS5}, we obtain induced metric on the worldsheet:
\begin{align}
d\bar s^2_{\rm WS} = \frac{1}{z^2}\frac{f(z) + \beta^2 z^4/z_h^4}{f^2(z)} dz^2 - \frac{2\beta^2}{z_h^2f(z)} dzdt - \frac{1}{z^2} [f(z) - \beta^2]dt^2. \label{hmetric_zt}
\end{align}
Apparently, an event horizon is developed on the worldsheet at $z=z_h(1-\beta^2)^{1/4}$, which is different from event horizon of target space \eqref{target_AdS5} once $\beta \neq 0$. In order to diagonalize the worldsheet metric \eqref{hmetric_zt}, we turn to the ``hatted'' coordinate system \cite{CasalderreySolana:2009rm}:
\begin{align}
& \hat{t}=\frac{t+\zeta_1(z)}{ \sqrt{\gamma} }, \qquad \qquad \hat{z}=\sqrt{\gamma}z, \qquad \qquad \hat{X}^i=\sqrt{\gamma} X^i, \nonumber \\
&\zeta_1(z)\equiv\frac{1}{2\pi T}\left[ \tan^{-1}(z\pi T)-\tanh^{-1}(z \pi T) \right] \nonumber  \\
&\qquad \quad -\frac{\sqrt{\gamma}}{2\pi T}\left[ \tan^{-1}(z \sqrt{\gamma}\pi T)-\frac{ 1 }{ 2 }\log \frac{ 1+z\sqrt{\gamma}\pi T}{ 1-z\sqrt{\gamma}\pi T }\right], \label{hat_X}
\end{align}
which brings the worldsheet metric \eqref{hmetric_zt} to AdS type:
\begin{align}
d\bar s_{\rm WS}^2 = \frac{1}{{\hat z}^2} \left[ \frac{d\hat z^2}{f(\hat z)} - f(\hat z) d\hat t^2 \right]. \label{Schw_metric}
\end{align}
Here, the Lorentz boost factor $\gamma = 1/\sqrt{1-\beta^2}$.

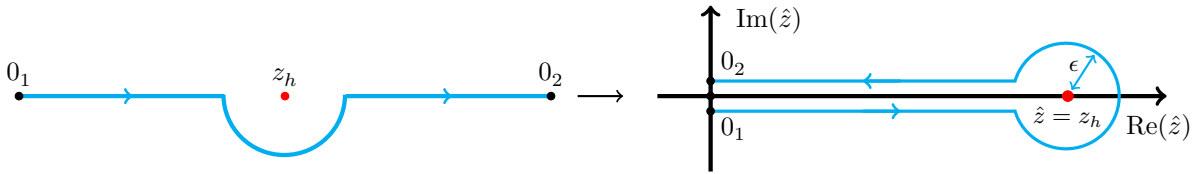
\begin{figure} [htbp!]
		\centering
		\begin{tikzpicture}[]
		
		\draw[cyan, ultra thick] (-5.5,0)--(-2.8,0);
		\draw[cyan, ultra thick] (-1.2,0)--(1.5,0);
		\draw[cyan, ultra thick] (-2.81,0.019) arc (-180:0:0.8);
		\draw[cyan, ->,very thick] (-1,0)--(0.2,0);
		\draw[cyan, ->,very thick] (-4.2,0)--(-4,0);
		\draw[fill] (-5.5,0) circle [radius=0.05];
		\node[above] at (-5.5,0) {\small $0_1$};
		\draw[fill] (1.5,0) circle [radius=0.05];
		\node[above] at (1.5,0) {\small $0_2$};
		\draw[fill,red] (-2,0) circle [radius=0.05];
		\node[above] at (-2,0) {\small $z_h$};				
		
		\draw[-to,thick] (1.85,0)--(2.45,0);

		\node[below] at (9.5,-0.1) {\small Re$(\hat{z})$};
		\draw[->,ultra thick] (2.9,0)--(9.6,0);
		\node[right] at (3.8,1) {\small Im$(\hat{z})$};
		\draw[->,ultra thick] (3.6,-1)--(3.6,1.2);
		\draw[cyan, very thick] (7.6,-0.18) arc (-165:165:0.7);
		\draw[cyan, very thick] (3.6,0.2)--(7.6,0.2);
		\draw[cyan, very thick] (3.6,-0.2)--(7.6,-0.2);
		\draw[cyan, ->,very thick] (5.6,-0.2)--(6.1,-0.2);
		\draw[cyan, <-,very thick] (5.6,0.2)--(6.1,0.2);
		\draw[fill] (3.6,0) circle [radius=0.05];
		\draw[fill] (3.6,0.2) circle [radius=0.05];
		\node[below] at (3.9,-0.15) {\small $0_1 $};
		\draw[fill] (3.6,-0.2) circle [radius=0.05];
		\node[above] at (3.9,0.15) {\small $0_2 $};
		\draw[fill , red ] (8.3,0) circle [radius=0.07];
		\node[below] at (8.3,0) {\footnotesize $\hat z=z_h $};
		\draw[cyan, thick,<->] (8.35,0.08)--(8.65,0.55);
		\node[above] at (8.38, 0.2) {\footnotesize $\epsilon$};
		\end{tikzpicture}
		\caption{From complexified (analytically continued near horizon) double AdS (left) \cite{Crossley:2015tka} to the holographic SK contour (right) \cite{Glorioso:2018mmw}. Indeed, the two horizontal legs overlap with the real axis.} \label{zcontour}
\end{figure}

In the ingoing EF coordinate system $\sigma^a= (\hat z, \hat v)$ with $d\hat v = d\hat t - d \hat z/f(\hat z) $, the worldsheet metric is free of coordinate singularity:
\begin{align}
d\bar s^2_{\rm WS} = \frac{1}{\hat z^2} \left[-2 d\hat v d\hat z-f(\hat z) d\hat v^2 \right]. \label{EF_metric}
\end{align}
We will be interesting in the fluctuations caused by Hawking radiation emitted from the worldsheet event horizon. Under the philosophy of \cite{Glorioso:2018mmw}, we double the worldsheet AdS, and analytically continue it around the emergent event horizon $\hat z =z_h$. Resultantly, the radial coordinate $\hat z$ varies along the contour of Figure \ref{zcontour}. This holographic contour forms an ideal prescription for studying effective action for a non-equilibrium problem.
The relationship between $\hat t$ and $\hat v$ is summarized below,
\begin{align}
\hat{v}&= \hat{t}+\chi_2(\hat z), \qquad \qquad \chi_2(\hat{z})\equiv - \int_{0_2}^{\hat{z}}\frac{ dy }{ f(y) }, \qquad \hat z\in [0_2, z_h + \epsilon], \nonumber \\
\hat{v}&= \hat{t}+\chi_1(\hat z), \qquad \qquad \chi_1(\hat{z})\equiv - \int_{0_1}^{\hat{z}}\frac{ dy }{ f(y) }, \qquad \hat z\in [0_1, z_h + \epsilon], \label{v_t_relation}
\end{align}
where the integration constants are fixed so that $\hat v$ and $\hat t$ equal on both AdS boundaries.

\subsection{String perturbation} \label{string_perturbation}

In order to study stochastic motion of the heavy quark, we turn on string perturbation on top of the trailing string background reviewed in subsection \ref{review_trailing_string}. 
%
%
Without loss of generality, we turn on string perturbation in the following way,
\begin{align}
& \hat X^1(\hat z, \hat v)=\sqrt{\gamma}\beta \left[ \sqrt \gamma \hat{t}(\hat v) -\zeta_1 (\hat{z}) + \zeta(\hat{z})\right] + \hat{X}_L(\hat z, \hat v),\nonumber \\
& \hat X^2(\hat z, \hat v)= 0 + \hat{X}_T (\hat z, \hat v), \qquad \qquad  \hat X^3(\hat z, \hat v)=0,
\end{align}
where we only need the relation $\hat t (\hat v)$ in its differential form.
Here, the string perturbations $\hat{X}_L$ and $\hat{X}_T$ are small in comparison with their backgrounds, so that the Nambu-Goto action \eqref{S_NG} could be expanded in powers of $\hat{X}_L$ and $\hat{X}_T$:
\begin{align}
S_{\rm NG} = - \frac{1}{2\pi \alpha^\prime} \int d\hat v \int_{0_1}^{0_2} d\hat z \sqrt{- \hat h(\hat X)} = S_{\rm NG}^{(0)} + S_{\rm NG}^{(1)} + S_{\rm NG}^{(2)} + S_{\rm NG}^{(3)} + \cdots, \label{S_NG_expansion}
\end{align}
where the radial integral is along the $\hat z$-contour of Figure \ref{zcontour} in a counter-clockwise way.
Note that $\hat X^i(\hat z \to 0)/\sqrt{\gamma}$ is the position of string endpoint. This shall be reflected in the AdS boundary conditions for $X_T, X_L$ to be presented later.

The zeroth order action vanishes since the integrand does not have any singularity inside the radial contour:
\begin{align}
S_{\rm NG}^{(0)}= -\frac{1}{2\pi \alpha^\prime} \int d\hat v \int_{0_1}^{0_2} \frac{d\hat z}{\hat z^2} =0. \label{S_NG_0th}
\end{align}

The next two terms in the expansion \eqref{S_NG_expansion}, say $S_{\rm NG}^{(1)}$ and $S_{\rm NG}^{(2)}$, were previously discussed in \cite{CasalderreySolana:2009rm} in the $(\hat z, \hat t)$-coordinate system. Here, for completeness we record them in the hatted ingoing EF coordinate system. The first order action $S_{\rm NG}^{(1)}$ is actually a boundary term:
\begin{align}
S_{\rm NG}^{(1)}=&\frac{1}{ 2\pi \alpha^\prime }\int d\hat{v} \int_{0_1}^{0_2} d\hat{z} \, \beta \gamma (\pi T)^2 \partial_{ \hat{z}} \hat{X}_L \nonumber \\
=& \int d\hat{v} \int_{0_1}^{0_2} d\hat{z}\, E \partial_{\hat{z}} \hat{X}_L= \int d\hat{v}  E \hat{X}_L\bigg|_{0_1}^{0_2}, \label{S_NG_1st}
\end{align}
which is exactly cancelled by the first order counterpart of boundary action $S_{\rm bdy}$ \eqref{S_bdy}. Here, we made use of the relation \eqref{beta-E}.

The quadratic order action $S_{\rm NG}^{(2)}$ is
\begin{align}
S_{\rm NG}^{(2)}=&-\frac{1}{ 2 \pi \alpha^{\prime} }\int{d\hat{v}}\int_{0_1}^{0_2}{d \hat{z}}\frac{ 1 }{ \hat{z}^2 } \left[-\gamma^2 \partial_{\hat{v}} \hat{X}_L \partial_{\hat{z}} \hat{X}_L +\frac{ \gamma^2 }{ 2 }f(\hat{z})(\partial_{\hat{z}} \hat{X}_L)^2 \right. \nonumber \\
&\qquad \qquad \qquad \qquad \qquad \quad \left. -\partial_{\hat{v}} \hat{X}_T \partial_{\hat{z}} \hat{X}_T +\frac{ 1 }{ 2 }f(\hat{z})(\partial_{\hat{z}} \hat{X}_T)^2\right]. \label{S_NG_2nd}
\end{align}
Apparently, the transverse dynamics behaves the same as the static case \cite{Son:2009vu}, while the longitudinal dynamics gets contracted by the Lorentz boost factor $\gamma$.

The third order action $S_{\rm NG}^{(3)}$ will be the focus of present work,
%
%
%
%
%
%
%
%
%
%
\begin{align}
S_{\rm NG}^{(3)}=&-\frac{\beta \gamma}{ 2 \pi \alpha^\prime }\int{d\hat{v}}\int_{0_1}^{0_2}d \hat{z} \left\{ \frac{\gamma^2 \pi^2 T^2}{2} f(\hat z) (\partial_{\hat z} \hat X_L)^3 + \frac{\pi^2 T^2}{2} f(\hat z) \partial_{\hat z} \hat X_L (\partial_{\hat z} \hat X_T)^2 \right. \nonumber \\
&\left. -\frac{\gamma^2}{2\hat z^2} (-1+3\pi^2 T^2 \hat z^2) (\partial_{\hat z} \hat X_L)^2 \partial_{\hat v} \hat X_L - \frac{1}{2\hat z^2} (1+\pi^2 T^2 \hat z^2) (\partial_{\hat z} \hat X_T)^2 \partial_{\hat v} \hat X_L \right. \nonumber \\
& \left. + \frac{ f(\hat z)}{\hat z^2 (1+\pi^2 T^2 \hat z^2)} \partial_{\hat z} \hat X_T \partial_{\hat z} \hat X_L \partial_{\hat v} \hat X_T - \frac{\gamma^2 }{\hat{z}^2[1+ (\hat{z}\pi T)^2 ]} \partial_{\hat{z}} \hat{X}_L (\partial_{\hat{v}} \hat{X}_L)^2 \right. \nonumber \\
& \left. - \frac{1}{\hat z^2 (1+\pi^2 T^2 \hat z^2)} \partial_{\hat z} \hat X_L (\partial_{\hat v} \hat X_T)^2\right\}. \label{S_NG_3rd}
\end{align}

In order to analyze time-reversal symmetry for the bulk theory, we turn to Schwarzschild coordinate system $\sigma^a = (\hat z, \hat t)$ momentarily, in which the Nambu-Goto action $S_{\rm NG}^{(2)}$ and $S_{\rm NG}^{(3)}$ are summarized in appendix \ref{S_NG_2nd_3rd_Schw}. From the worldsheet perspective, the second order action $S_{\rm NG}^{(2)}$ is invariant under $\mathcal{T}$-transformation: $\hat t \to - \hat t, ~ \hat X_{L,T}(\hat t, \hat z) \to \hat X_{L,T}(-\hat t, \hat z)$. However, the cubic order action $S_{\rm NG}^{(3)}$ explicitly breaks $\mathcal{T}$-symmetry. $\mathcal T$-symmetry is crucial in deriving KMS constraints on the SK effective action \cite{Crossley:2015evo}, which will be discussed in section \ref{SK_effetive_action}.

%

The EOMs for $\hat X_{T,L}$, derived from \eqref{S_NG_2nd} and \eqref{S_NG_3rd}, are
\begin{align}
&0=\partial_{\hat z} \left[ \frac{1}{\hat z^2} f(\hat z) \partial_{\hat z} \hat X_T \right] - \frac{2}{\hat z^2} \partial_{\hat v} \partial_{\hat z} \hat X_T + \frac{2}{\hat z^3} \partial_{\hat v} \hat X_T + f_T[\hat X_T, \hat X_L], \nonumber \\
&0=\partial_{\hat z} \left[ \frac{1}{\hat z^2} f(\hat z) \partial_{\hat z} \hat X_L \right] - \frac{2}{\hat z^2} \partial_{\hat v} \partial_{\hat z} \hat X_L + \frac{2}{\hat z^3} \partial_{\hat v} \hat X_L + f_L[\hat X_T, \hat X_L], \label{eom_X_TL}
\end{align}
where the nonlinear terms $f_T, f_L$ are quadratic in $\hat X_T, \hat X_L$. We do not record explicit expressions for $f_T, f_L$ since they are irrelevant for the computation of cubic order SK effective action. We will elaborate on this point later. The nonlinear EOMs \eqref{eom_X_TL} will be solved perturbatively
\begin{align}
\hat X_{T,L} = \alpha^1 \hat X_{T,L}^{(1)} + \alpha^2 \hat X_{T,L}^{(2)} + \cdots. \label{X_expansion}
\end{align}
Then, at each order in the perturbative expansion \eqref{X_expansion}, $\hat X_{T,L}^{(n)}$ obeys linear differential equations. Particularly, at the lowest order $\mathcal{O}(\alpha^1)$, $\hat X_{T,L}$ satisfy the same homogeneous equation:
\begin{align}
\partial_{\hat z} \left[ \frac{1}{\hat z^2} f(\hat z) \partial_{\hat z} \Phi \right] - \frac{2}{\hat z^2} \partial_{\hat v} \partial_{\hat z} \Phi + \frac{2}{\hat z^3} \partial_{\hat v} \Phi=0, \qquad \Phi = \hat X_T^{(1)} ~~ {\rm or}~~ \hat X_L^{(1)}, \label{eom_X_TL_linearized}
\end{align}
which is analogous to EOM for a free scalar in Schwarzschild-AdS$_2$. The AdS boundary conditions are
\begin{align}
& \hat X_T^{(1)} (\hat z \to 0_s) = \hat q_s^T, \qquad \qquad  \hat X_T^{(n>1)} (\hat z \to 0_s) =0, \qquad s=1,2, \nonumber \\
& \hat X_L^{(1)} (\hat z \to 0_s) = \hat q_s^L, \qquad \qquad  \hat X_L^{(n>1)} (\hat z \to 0_s) =0, \qquad s=1,2. \label{AdS_bdy_condition}
\end{align}

Substituting the expansion \eqref{X_expansion} into \eqref{S_NG_2nd}, the quadratic order action $S_{\rm NG}^{(2)}$ can be split into two parts:
\begin{align}
S_{\rm NG}^{(2)} = S_{\rm NG}^{(2), \rm P1} + S_{\rm NG}^{(2), \rm P2}, \label{S_NG_2nd_expand}
\end{align}
where
\begin{align}
S_{\rm NG}^{(2), \rm P1}=&-\frac{1}{2\pi \alpha^\prime}\int d\hat{v}\int_{0_1}^{0_2} d \hat{z} \frac{1}{\hat{z}^2} \left[-\gamma^2 \partial_{\hat{v}} \hat{X}^{(1)} _L \partial_{\hat{z}} \hat{X}^{(1)} _L +\frac{ \gamma^2 }{ 2 }f(\hat{z})(\partial_{\hat{z}} \hat{X}^{(1)} _L)^2 \right.\nonumber \\
&\qquad \qquad \qquad \qquad \qquad \quad \left. -\partial_{\hat{v}} \hat{X}^{(1)} _T \partial_{\hat{z}} \hat{X}^{(1)} _T +\frac{ 1 }{ 2 }f(\hat{z})(\partial_{\hat{z}} \hat{X}^{(1)} _T)^2\right], \label{S_NG_2nd_P1}
\end{align}
\begin{align}
S_{\rm NG}^{(2), \rm P2}= & -\frac{1}{ 2 \pi \alpha^\prime }\int d\hat{v} \int_{0_1}^{0_2} d\hat{z} \frac{1} {\hat{z}^2} \left[-\partial_{\hat{v}}X_T^{(2)}\partial_{\hat{z}}\hat{X}_T^{(1)}-\partial_{\hat{v}}
\hat{X}_T^{(1)}\partial_{\hat{z}}\hat{X}_T^{(2)}+ f(\hat{z}) \partial_{\hat{z}} \hat{X}_T^{(2)}\partial_{\hat{z}}\hat{X}_T^{(1)} \right. \nonumber \\
& \qquad \left.+ \gamma^2 \left(-\partial_{\hat{v}}\hat{X}_L^{(2)} \partial_{\hat{z}} \hat{X}_L^{(1)}-\partial_{\hat{v}}\hat{X}_L^{(1)}\partial_{\hat{z}}\hat{X}_L^{(2)} + f (\hat{z}) \partial_{\hat{z}} \hat{X}_L^{(2)} \partial_{\hat{z}} \hat{X}_L^{(1)} \right) \right]. \label{S_NG_2nd_P2}
\end{align}
In \eqref{S_NG_2nd_expand}, a piece of order $\mathcal{O}(\alpha^4)$ is ignored since it is beyond scope of present work.
Upon integration by parts, \eqref{S_NG_2nd_P2} is computed as
\begin{align}
S_{\rm NG}^{(2), \rm P2}=& - \frac{1}{2\pi \alpha^\prime} \int d\hat v \int_{0_1}^{0_2} d\hat z \left\{ \partial_{\hat z} \left( \hat X_T^{(2)} \frac{f(\hat z)}{\hat z^2} \partial_{\hat z} \hat X_T^{(1)} - \hat X_T^{(2)} \frac{1}{\hat z^2} \partial_{\hat v} \hat X_T^{(1)}\right) \right. \nonumber \\
& \qquad \qquad \qquad \qquad \qquad \left. +\gamma^2 \partial_{\hat z} \left( \hat X_L^{(2)} \frac{f(\hat z)}{\hat z^2} \partial_{\hat z} \hat X_L^{(1)} - \hat X_L^{(2)} \frac{1}{\hat z^2} \partial_{\hat v} \hat X_L^{(1)}\right) \right\}, \nonumber \\
= & - \frac{1}{2\pi \alpha^\prime} \int d\hat v \left\{ \hat X_T^{(2)} \left(  \frac{f(\hat z)}{\hat z^2} \partial_{\hat z} \hat X_T^{(1)} - \frac{1}{\hat z^2} \partial_{\hat v} \hat X_T^{(1)}\right) \right. \nonumber \\
& \qquad \qquad \qquad \quad \left. +\gamma^2 \hat X_L^{(2)} \left( \frac{f(\hat z)}{\hat z^2} \partial_{\hat z} \hat X_L^{(1)} - \frac{1}{\hat z^2} \partial_{\hat v} \hat X_L^{(1)}\right) \right\}\bigg|_{\hat z= 0_1}^{\hat z= 0_2} \nonumber \\
= & 0, \label{S_NG_2nd_P2=0}
\end{align}
where the Dirichlet conditions for $\hat X_{T,L}^{(2)}$ at the AdS boundaries have been utilized, cf. \eqref{AdS_bdy_condition}. Moreover, in obtaining \eqref{S_NG_2nd_P2=0}, total time derivative terms are ignored. In the same way, \eqref{S_NG_2nd_P1} is cast into a surface term:
\begin{align}
S_{\rm NG}^{(2), \rm P1}= -\frac{1}{2\pi \alpha^\prime} \int d \hat v \left[ \frac{f(\hat z)}{2 \hat z^2} \hat X_T^{(1)} \partial_{\hat z} \hat X_T^{(1)} + \gamma^2 \frac{f(\hat z)}{2 \hat z^2} \hat X_L^{(1)} \partial_{\hat z} \hat X_L^{(1)} \right]\bigg|_{\hat z =0_1}^{\hat z = 0_2}, \label{S_NG_2nd_P1_bdy}
\end{align}
where total time derivative terms are also ignored.

Therefore, accurate to cubic order $\mathcal{O}(\alpha^3)$, the total string action (Nambu-Goto action plus the boundary term) becomes
\begin{align}
S=S_{\rm NG}+ S_{\rm bdy}= S_{\rm NG}^{(2), \rm P1} + S_{\rm NG}^{(3)}\big|_{\hat X_{T,L} \to \hat X_{T,L}^{(1)}}. \label{S_NG+bdy_cubic}
\end{align}
Thus, in order to derive boundary effective action up to cubic order in the position $\hat q^{T,L}$, we just need to solve the linearized EOMs \eqref{eom_X_TL_linearized} for string perturbations. This is similar to the case of a {\it static} heavy quark in plasma medium \cite{Chakrabarty:2019aeu}. Once the linearized EOMs \eqref{eom_X_TL_linearized} are solved, the core task will be to compute the cubic action $S_{\rm NG}^{(3)}\big|_{\hat X_{T,L} \to \hat X_{T,L}^{(1)}}$, which involves a number of radial contour integrals.

\section{Solution for string fluctuation on the contour} \label{linearized_string_solution}

In this section, we solve the linearized EOMs \eqref{eom_X_TL_linearized} over the radial contour of Figure \ref{zcontour}. The solutions were previously obtained in \cite{Chakrabarty:2019aeu} through heuristic analysis with the help of time-reversal symmetry. Here, we elaborate on the derivation for completeness, following the idea of \cite{Bu:2020jfo}. To this end, we cut the contour in Figure \ref{zcontour} at $\hat z = \hat z_{\pm} \equiv \hat z_h + \epsilon$, where the subscripts ``$_+$'' and ``$_-$'' are used to distinguish between the upper and lower segments of the contour. Then, we will search for general solutions for \eqref{eom_X_TL_linearized} in the single copy of the contour, namely, either on the upper branch or on the lower branch of Figure \ref{zcontour}. Crucially, the general solution for the upper branch and that for the lower branch have to be carefully glued at the cutting point $\hat z = \hat z_h + \epsilon$. The correct gluing conditions at the cutting point $\hat z = \hat z_h + \epsilon$ will be derived from the quadratic order action \eqref{S_NG_2nd}.

\subsection{Horizon gluing conditions}

Since the actions for $\hat X_T^{(1)}$ and $\hat X_L^{(1)}$ differ by an overall constant $\gamma^2$, the horizon gluing conditions for $\hat X_T^{(1)}$ and $\hat X_L^{(1)}$ shall be the same. So, we will take $\hat X_T^{(1)}$ as an example to derive the horizon gluing conditions. The action for $\hat X_T^{(1)}$ is
\begin{align}
S_{\rm NG}^{(2),\rm T}=-\frac{1}{ 2 \pi \alpha^\prime }\int{d\hat{v}} \int_{0_1}^{0_2} {d \hat{z}}\frac{ 1 }{ \hat{z}^2 } \left[-\partial_{\hat{v}} \hat{X}_T^{(1)}  \partial_{\hat{z}} \hat{X}^{(1)}_T  +\frac{ 1 }{ 2 }f(\hat{z})\left(\partial_{\hat{z}} \hat{X}^{(1)}_T \right)^2\right]. \label{S_NG_2nd_T}
\end{align}
In the spirit of \cite{Skenderis:2008dg}, the gluing conditions at the cutting point can be derived by demanding the bulk action $S_{\rm NG}^{(2),\rm T}$ to be extremal with respect to variation of horizon data $\hat X_T^{(1)}(\hat z_h + \epsilon, \hat v)$. The variation of \eqref{S_NG_2nd_T} is
\begin{align}
\delta S_{\rm NG}^{(2),\rm T} =& -\frac{1}{2\pi\alpha^\prime} \int d\hat{v} \left\{\left(-\frac{ 1 }{\hat{z}^2}\partial_{ \hat{v}} \hat{X}_T^{(1)}+ \frac{f(\hat z)} {\hat{z}^2} \partial_{ \hat{z}} \hat{X}_T^{(1)}\right) \delta \hat{X}_T^{(1)} \bigg|_{0_1}^{\hat{z}_-} \right. \nonumber \\
& \qquad \qquad \qquad \quad \left. - \left(-\frac{ 1 }{\hat{z}^2  }\partial_{ \hat{v}} \hat{X}_T^{(1)} + \frac{f(\hat z)}{\hat{z}^2}\partial_{ \hat{z}} \hat{X}\right) \delta \hat{X}_T^{(1)} \bigg|_{0_2}^{\hat{z}_+} \right\},
\end{align}
where the linearized EOM for $\hat X_T^{(1)}$ (cf.~\eqref{eom_X_TL_linearized}) is used.
It is natural to assume that $\hat X_T^{(1)}$ is continuous across the cutting point:
\begin{align}
\hat X_T^{(1)}(\hat z_+) = \hat X_T^{(1)}(\hat z_-). \label{X_T_continuity}
\end{align}
Then, the extremum condition at the cutting point gives rise to
\begin{align}
\frac{\delta S_{\rm NG}^{(2), \rm T}}{\delta \hat X_T^{(1)}(\hat z_h+\epsilon, \hat v)}=0 \Longrightarrow
f(\hat z) \partial_{\hat z} \hat X_T^{(1)}\big|_{\hat z= \hat z_+} = f(\hat z) \partial_{\hat z} \hat X_T^{(1)}\big|_{\hat z= \hat z_-}. \label{der_X_T_continuity}
\end{align}
Since $f(\hat z)$ vanishes at the horizon, the gluing condition implies $\partial_{\hat z} \hat X_T^{(1)}$ may not be continuous across the horizon.

The same analysis can be made for the longitudinal mode $\hat X_L^{(1)}$, which will give the same gluing conditions as \eqref{X_T_continuity} and \eqref{der_X_T_continuity}:
\begin{align}
\hat X_T^{(1)}(\hat z_+) = \hat X_T^{(1)}(\hat z_-), \qquad  f(\hat z) \partial_{\hat z} \hat X_L^{(1)}\big|_{\hat z= \hat z_+} = f(\hat z) \partial_{\hat z} \hat X_L^{(1)}\big|_{\hat z= \hat z_-}. \label{X_L+der_X_L_continuity}
\end{align}

Physically, the gluing conditions \eqref{X_T_continuity}, \eqref{der_X_T_continuity} and \eqref{X_L+der_X_L_continuity} correspond to continuity for string's momentum. We will use them to glue the piecewise solutions to be found for the upper segment and the lower segment in next two subsections.

\subsection{From linearly independent solutions to solution on the contour}

In this subsection, we search for linearly independent solutions for $\hat X_{T, L}^{(1)}$ on upper or lower segments of the contour. Then, the general solution (superposition of linearly independent solutions) will be glued at the cutting point, under the conditions \eqref{X_T_continuity}, \eqref{der_X_T_continuity} and \eqref{X_L+der_X_L_continuity}. Eventually, the AdS boundary conditions \eqref{AdS_bdy_condition} will determine all the integration constants.

It is more convenient to turn to frequency domain
\begin{align}
\Phi(\hat z, \hat v) = \int \frac{d\hat\omega}{2\pi} e^{-i\hat \omega \hat v} \Phi(\hat z, \hat \omega),
\end{align}
where the Fourier mode $\Phi(\hat z, \hat \omega)$ satisfies a homogeneous ordinary differential equation (ODE):
\begin{align}
0=\partial_{\hat{z}}\left[\frac{f(\hat z)}{\hat{z}^2}\partial_{\hat{z}} \Phi \right] +\frac{2}{\hat{z}^2} i \hat \omega \partial_{\hat{z} }\Phi-\frac{2}{\hat{z}^3} i \hat \omega \Phi. \label{eom_Phi}
\end{align}
Apparently, the time-reversal symmetry is not realized simply as
\begin{align}
\hat \omega \to -\hat \omega, \qquad \Phi(\hat z, \hat \omega) \to \Phi(\hat z, - \hat \omega).
\end{align}
In order to see how the time-reversal symmetry is realized, we temporarily turn to Schwarzschild coordinate $(\hat z, \hat t)$ for the worldsheet space, in which the linearized string fluctuations $\hat X_{T,L}^{(1)}(\hat z, \hat t)$ will be collectively denoted as $\tilde \Phi(\hat z, \hat t)$. In the frequency domain:
\begin{align}
\tilde \Phi(\hat z, \hat t) = \int \frac{d\hat\omega}{2\pi} e^{-i\omega \hat t} \tilde \Phi(\hat z, \omega),
\end{align}
where the Fourier mode $\tilde \Phi(\hat z, \hat\omega)$ obeys
\begin{align}
0=\partial_{\hat{z}}\left[\frac{f(\hat z)}{\hat{z}^2} \partial_{\hat{z}} \tilde \Phi \right] + \frac{\hat\omega^2}{\hat{z}^2 f(\hat z)} \tilde \Phi. \label{eom_Phi_tilde}
\end{align}
Here, time-reversal symmetry is simply realized as
\begin{align}
\hat\omega \to -\hat\omega, \qquad \tilde\Phi(\hat z, \hat\omega) \to \tilde \Phi(\hat z, -\hat\omega), \label{T-symmetry}
\end{align}
which will be helpful for finding linearly independent solutions.

In the single copy of the contour, namely, $\hat z \in [0,z_h]$, the EOM \eqref{eom_Phi_tilde} admits two linearly independent solutions: the ingoing mode and outgoing mode. Near the horizon, the ingoing mode behaves as
\begin{align}
\tilde \Phi^{\rm ig}(\hat z \to z_h,\hat\omega) = (z_h - \hat z)^{-i\hat\omega/(4\pi T)} \left[\tilde \Phi_h^0 + \tilde \Phi_h^1 (z_h - \hat z) + \tilde \Phi_h^2 (z_h - \hat z)^2 + \cdots  \right], \label{Phi_tilde_horizon}
\end{align}
where $\tilde \Phi_h^1, \tilde \Phi_h^2, \cdots$ are determined in terms of $\tilde \Phi_h^0$, and the horizon data $\tilde \Phi_h^0$ is undetermined. The ingoing solution $\tilde \Phi^{\rm ig}$ is uniquely fixed once the horizon data $\tilde \Phi_h^0$ is given. Indeed, the horizon data $\tilde \Phi_h^0$ is in one-to-one correspondence to the boundary value $\tilde \Phi^{\rm ig}(\hat z=0, \hat\omega)$.

Under the time-reversal symmetry \eqref{T-symmetry}, the outgoing mode is
\begin{align}
\tilde \Phi^{\rm og}(\hat z,\hat\omega) = \tilde \Phi^{\rm ig}(\hat z, -\hat\omega). \label{ig_og_Schw}
\end{align}
Now, we would like to convert the relationship \eqref{ig_og_Schw} into the ingoing EF coordinate. This is achieved by realizing that $\Phi$ (also $\tilde\Phi$) is a scalar in the worldsheet space, which tells that
\begin{align}
\Phi(\hat z, \hat\omega) e^{-i\hat\omega \hat v} = \tilde \Phi(\hat z, \hat\omega) e^{-i\hat\omega \hat t} \Longrightarrow \Phi(\hat z, \hat\omega) = \tilde \Phi(\hat z, \hat\omega) e^{i\hat\omega \chi_s(\hat z)}, \qquad s=1,2, \label{Phi_tilde_Phi}
\end{align}
where the subscript ``$s$'' is to remind the upper or lower segment. Here, we made use of the relation \eqref{v_t_relation}. Notice that the relationship \eqref{Phi_tilde_Phi} also holds for each linearly independent solution. So, in the ingoing EF coordinate, an analogue of \eqref{ig_og_Schw} is
\begin{align}
\Phi^{\rm og}(\hat z, \hat\omega) = e^{2i\hat\omega \chi_s(\hat z)} \Phi^{\rm ig}(\hat z, -\hat\omega),
\end{align}
which could interpreted as nonlinear realization of time-reversal symmetry \cite{Chakrabarty:2019aeu}.

Thus, we immediately write down general solutions on the upper and lower segments:
\begin{align}
\Phi^{\rm up}(\hat z, \hat\omega)& = c^{\rm up} \Phi^{\rm ig}(\hat z, \hat\omega) + d^{\rm up} e^{2i\hat\omega \chi_2(\hat z)} \Phi^{\rm ig}(\hat z, -\hat\omega), \,\,\qquad \hat z \in [0_2, z_h+\epsilon], \nonumber \\
\Phi^{\rm dw}(\hat z, \hat\omega)& = c^{\rm dw} \Phi^{\rm ig}(\hat z, \hat\omega) + d^{\rm dw} e^{2i\hat\omega \chi_1(\hat z)} \Phi^{\rm ig}(\hat z, -\hat\omega), \qquad \hat z \in [0_1, z_h+\epsilon]. \label{Phi_up_dw1}
\end{align}
Here, since the ingoing solution $\Phi^{\rm ig}$ is regular, we do not need to distinguish between the upper and lower segments. Imposing the gluing conditions \eqref{X_T_continuity}, \eqref{der_X_T_continuity} and \eqref{X_L+der_X_L_continuity} on the solutions \eqref{Phi_up_dw1}, we obtain
\begin{align}
c^{\rm up} = c^{\rm dw}, \qquad d^{\rm up} = d^{\rm dw} e^{\beta_0 \hat\omega},
\end{align}
where $\beta_0$ is the inverse temperature $1/T$.
So, the general solutions \eqref{Phi_up_dw1} become
\begin{align}
\Phi^{\rm up}(\hat z, \hat\omega)& = c \Phi^{\rm ig}(\hat z, \hat\omega) + d e^{2i\hat\omega \chi_2(\hat z)} \Phi^{\rm ig}(\hat z, -\hat\omega), \qquad \qquad \hat z \in [0_2, z_h+\epsilon], \nonumber \\
\Phi^{\rm dw}(\hat z, \hat\omega)& = c \Phi^{\rm ig}(\hat z, \hat\omega) + d e^{-\beta\hat\omega} e^{2i\hat\omega \chi_1(\hat z)} \Phi^{\rm ig}(\hat z, -\hat\omega), \qquad \hat z \in [0_1, z_h+\epsilon], \label{Phi_up_dw2}
\end{align}
where a relabelling $c^{\rm up} \to c, d^{\rm up} \to d$ was made. The piecewise solutions \eqref{Phi_up_dw2} could be put into a compact form:
\begin{align}
\Phi(\hat z, \hat\omega)& = c \Phi^{\rm ig}(\hat z, \hat\omega) + d e^{2i\hat\omega \chi(\hat z)} \Phi^{\rm ig}(\hat z, -\hat\omega), \qquad \qquad \hat z \in [0_1, 0_2],
\end{align}
where the function $\chi(\hat z)$ is defined over the entire contour:
\begin{align}
\chi(\hat z) \equiv - \int_{0_2}^{\hat z} \frac{dy}{f(y)}, \qquad \hat z \in [0_1,0_2]. \label{chi_on_contour}
\end{align}
The integration constants $c,d$ are fixed by the AdS boundary conditions:
\begin{align}
& c \Phi^{\rm ig(0)}(\hat\omega) + d \Phi^{\rm ig(0)}(-\hat\omega)= \hat q_2(\hat\omega), \qquad
c \Phi^{\rm ig(0)}(\hat\omega) + d e^{-\beta_0\hat\omega} \Phi^{\rm ig(0)}(-\hat\omega) = \hat q_1(\hat\omega) \nonumber \\
\Rightarrow & c= \frac{1}{2}\coth\frac{\beta_0 \hat\omega}{2} \frac{\hat q_a(\hat\omega)} {\Phi^{\rm ig(0)}(\hat\omega)} + \frac{\hat q_r(\hat\omega)}{\Phi^{\rm ig(0)}(\hat\omega)}, \qquad d= - \frac{\hat q_a(\hat\omega)}{(1-e^{-\beta_0 \hat\omega})\Phi^{\rm ig(0)}(-\hat\omega)},
\end{align}
where $\Phi^{\rm ig(0)}(\hat\omega) = \Phi^{\rm ig}(\hat z=0, \hat\omega)$. Here, we introduced the $(r,a)$-basis:
\begin{align}
\hat q_r = \frac{1}{2}(\hat q_1 + \hat q_2), \qquad \qquad \hat q_a = \hat q_1 - \hat q_2.
\end{align}
Recovering the superscripts ``$T, L$'' in the position $\hat q_r, \hat q_a$, we obtain explicit solutions for linearized string perturbation $\hat X_{T,L}^{(1)}$ over the entire contour:
\begin{align}
\hat X_{T,L}^{(1)}(\hat z, \hat\omega) = A(\hat z, \hat\omega) \hat q_r^{T,L}(\hat\omega) + B(\hat z, \hat\omega) \hat q_a^{T,L}(\hat\omega), \qquad \hat z \in [0_1, 0_2], \label{X_TL_1}
\end{align}
where
\begin{align}
A(\hat z, \hat\omega) = \frac{\Phi^{\rm ig}(\hat z, \hat\omega)}{\Phi^{\rm ig(0)}(\hat\omega)}, \qquad B(\hat z, \hat\omega) = \frac{1}{2}\coth\frac{\beta_0\hat\omega}{2} \frac{\Phi^{\rm ig}(\hat z, \hat\omega)}{\Phi^{\rm ig(0)}(\hat\omega)} - \frac{e^{2i\hat\omega \chi(\hat z)}}{1-e^{-\beta_0 \hat\omega}} \frac{\Phi^{\rm ig}(\hat z, -\hat\omega)}{\Phi^{\rm ig(0)} (-\hat\omega)}. \label{A_B}
\end{align}
Thus, solving linearized string perturbation over the entire contour boils down to searching for ingoing solution $\Phi^{\rm ig}(\hat z, \hat\omega)$ in single copy of the radial contour.

%

\section{Schwinger-Keldysh effective action for Brownian particle} \label{SK_effetive_action}

In this section, we plug the solution for linearized string perturbation of section \ref{linearized_string_solution} into $S_{\rm NG}^{(2), \rm P1}$ \eqref{S_NG_2nd_P1_bdy} and $S_{\rm NG}^{(3)}$ \eqref{S_NG_3rd}, and evaluate the radial integral, producing the SK effective action for heavy quark:
\begin{align}
S_{eff}^{(2)} = S_{\rm NG}^{(2), \rm P1}, \qquad \qquad \qquad S_{eff}^{(3)} = S_{\rm NG}^{(3)}\big|_{\hat X_{T,L} \to \hat X^{(1)}_{T,L}}.
\end{align}
%

\subsection{Quadratic action: coloured noise}

In order to compute the quadratic order action \eqref{S_NG_2nd_P1_bdy}, we need near-boundary expansion for $e^{2i\hat\omega \chi_{1,2}(\hat z)}$:
\begin{align}
e^{2i\hat\omega \chi_s(\hat z)} \xrightarrow[]{\hat z \to 0_s} 1 -2i\hat\omega \hat z - 2\hat\omega^2 \hat z^2 + \frac{4}{3}i\hat\omega^3 \hat z^3 + \cdots, \qquad s=1~~{\rm or}~~2.
\end{align}
Near the AdS boundary, the ingoing mode $\Phi^{\rm ig}$ behaves as
\begin{align}
\Phi^{\rm ig}(\hat z \to 0, \hat\omega)= \Phi^{\rm ig(0)}(\hat\omega) -i\hat\omega \Phi^{\rm ig(0)}(\hat\omega) \hat z + \Phi^{\rm ig(3)}(\hat\omega) \hat z^3 + \cdots. \label{Phi_ig_bdy}
\end{align}
Then, the near-boundary behaviors of $\hat X_{T,L}^{(1)}$ are
\begin{align}
\hat X_{T,L}^{(1)}(\hat z \to 0_s, \hat\omega) = \hat q_s^{T,L}(\hat\omega) - i\hat\omega \hat q_s^{T,L}(\hat\omega) \hat z + \mathbb{O}^{T,L}_s(\hat\omega) \hat z^3 + \cdots, \qquad s= 1 ~~ {\rm or} ~~ 2, \label{X_TL_bdy}
\end{align}
where the normalizable modes $\mathbb{O}^{T,L}_s(\hat\omega)$ are expressed in terms of ratio $\Phi^{\rm ig(3)}/\Phi^{\rm ig(0)}$:
\begin{align}
\mathbb{O}^{T,L}_2(\hat\omega)= &\frac{\Phi^{\rm ig(3)}(\hat\omega)}{\Phi^{\rm ig(0)}(\hat\omega)} \hat q_r^{T,L}(\hat\omega) + \frac{1}{2}\coth\frac{\beta_0\hat\omega}{2} \frac{\Phi^{\rm ig(3)}(\hat\omega)}{\Phi^{\rm ig(0)}(\hat\omega)} \hat q_a^{T,L}(\hat\omega) \nonumber \\
& - \frac{1}{1-e^{-\beta_0\hat\omega}} \frac{\Phi^{\rm ig(3)}(-\hat\omega)}{\Phi^{\rm ig(0)}(-\hat\omega)} \hat q_a^{T,L}(\hat\omega) + \frac{2i\hat\omega^3}{3(1-e^{-\beta_0\hat\omega})} \hat q_a^{T,L}(\hat\omega), \nonumber \\
\mathbb{O}^{T,L}_1(\hat\omega)= &\frac{\Phi^{\rm ig(3)}(\hat\omega)}{\Phi^{\rm ig(0)}(\hat\omega)} \hat q_r^{T,L}(\hat\omega) + \frac{1}{2}\coth\frac{\beta_0\hat\omega}{2} \frac{\Phi^{\rm ig(3)}(\hat\omega)}{\Phi^{\rm ig(0)}(\hat\omega)} \hat q_a^{T,L}(\hat\omega) \nonumber \\
& - \frac{e^{-\beta_0\hat\omega}}{1-e^{-\beta_0\hat\omega}} \frac{\Phi^{\rm ig(3)}(-\hat\omega)} {\Phi^{\rm ig(0)}(-\hat\omega)} \hat q_a^{T,L}(\hat\omega) + \frac{2i\hat\omega^3 e^{-\beta_0\hat\omega}} {3(1-e^{-\beta_0\hat\omega})} \hat q_a^{T,L}(\hat\omega).
\end{align}
Plugging \eqref{X_TL_bdy} into \eqref{S_NG_2nd_P1_bdy}, we obtain the quadratic order SK effective action
%
%
\begin{align}
S_{eff}^{(2)}&= \frac{1}{2\pi \alpha^\prime}\int \frac{d\hat\omega}{2\pi} \left\{ \frac{i}{2} \hat q_a^T(-\hat\omega) \hat G_{rr}(\hat\omega) \hat q_a^T(\hat\omega) + \hat q_a^T(-\hat\omega) \left[ \hat M \hat\omega^2 + \hat G_{ra}(\hat\omega) \right]\hat q_r^T(\hat\omega) \right\} \nonumber \\
&+\frac{\gamma^2}{2\pi \alpha^\prime}\int \frac{d\hat\omega}{2\pi} \left\{\frac{i}{2}\hat q_a^L(-\hat\omega) \hat G_{rr}(\hat\omega) \hat q_a^L(\hat\omega) + \hat q_a^L(-\hat\omega) \left[ \hat M \hat\omega^2 + \hat G_{ra}(\hat\omega) \right] \hat q_r^L(\hat\omega) \right\}, \label{Seff_2nd}
\end{align}
where
\begin{align}
&\hat G_{rr}(\hat\omega)= -i\coth\frac{\beta_0\hat\omega}{2}\left[\frac{3}{2} \frac{\Phi^{\rm ig(3)}(\hat\omega)} {\Phi^{\rm ig(0)}(\hat\omega)} - \frac{3}{2} \frac{\Phi^{\rm ig(3)}(-\hat\omega)} {\Phi^{\rm ig(0)}(-\hat\omega)} + i\hat\omega^3 \right], \nonumber \\
&\hat G_{ra}(\hat\omega)=3\frac{\Phi^{\rm ig(3)}(\hat\omega)} {\Phi^{\rm ig(0)} (\hat\omega)} + i\hat\omega^3. \label{Grr_Gra}
\end{align}
The quadratic action \eqref{Seff_2nd} was previously obtained in \cite{Giecold:2009cg,CasalderreySolana:2009rm} by using Kruskal extension of the AdS/CFT correspondence \cite{Herzog:2002pc}. In terms of the rescaled variables $\hat q^{T, L}$, \eqref{Seff_2nd} looks essentially the same as that of static case \cite{Son:2009vu}. In the Schwarzschild coordinate, the $i\hat\omega^3$ piece in $\hat G_{rr}$ and $\hat G_{ra}$ will be absorbed into the normalizable mode $\tilde \Phi^{\rm ig(3)}$:
\begin{align}
\frac{\Phi^{\rm ig(3)}(\hat\omega)}{\Phi^{\rm ig(0)}(\hat\omega)} = \frac{\tilde \Phi^{\rm ig(3)}(\hat\omega)}{\tilde \Phi^{\rm ig(0)}(\hat\omega)} - \frac{1}{3}i\hat\omega,
\end{align}
which will render the correlators \eqref{Grr_Gra} to be of the same form as in \cite{Son:2009vu} (see (A11) of \cite{Son:2009vu}).
In \eqref{Seff_2nd}, the $\hat M \hat\omega^2$-term arises from the divergent part of the AdS boundary limit $\hat z\to 0$ \cite{Son:2009vu}: $\hat M = \lim_{\hat z \to 0} 1/{\hat z}$, which is related to zero-temperature quark mass $M=\lim_{z\to 0}1/z$ \cite{Herzog:2006gh} by the boost factor: $\hat M = M/\sqrt{\gamma}$. Apparently, the quark mass becomes finite if the quark lives on a finite cutoff slice $z= \Lambda \neq0$.

In SK effective action, the quark position $q$ couples to plasma medium force $\mathcal F$ in the form $q_1 \mathcal F_1 - q_2 \mathcal F_2 = q_r \mathcal F_a + q_a \mathcal F_r $ \cite{Son:2009vu}. Thus, $S_{eff}$ acts as generating functional $W$ for correlators of medium force $\mathcal F$:
\begin{align}
W[\hat q_r^{L,T}, \hat q_a^{L,T}] = i S_{eff}[\hat q_r^{L,T}, \hat q_a^{L,T}],
\end{align}
which, for convenience, is presented in terms of rescaled quantities. 
Then, we could interpret $\hat G_{rr}$ and $\hat G_{ra}$ as the symmetric and retarded force-force\footnote{Here, the force should be understood as rescaled one $\hat{\mathcal F} = \mathcal F/\sqrt{\gamma}.$} correlators \cite{Son:2009vu}. Obviously, the following relation holds
\begin{align}
\hat G_{rr}(\hat\omega)= \coth\frac{\beta_0\hat\omega}{2}{\rm Im} [\hat G_{ra} (\hat\omega)], \label{KMS_2point}
\end{align}
which is more easily understood from worldsheet perspective. The relation \eqref{KMS_2point} arises from the following fact: the state described by \eqref{EF_metric} is thermal, and the theory \eqref{S_NG_2nd} preserves $\mathcal T$-symmetry. In terms of physical quantities by $\hat \omega \to \omega, \hat q \to q, \hat {\mathcal F} \to \mathcal F$, \eqref{KMS_2point} will not take the exact form of KMS condition for a thermal state, which implies the fluctuation and dissipation felt by relativistic heavy quark is of non-thermal nature \cite{Giecold:2009cg}.

Via relabelling $\hat\omega \to -\hat\omega$ in the $ra$-term of \eqref{Seff_2nd}, we can read off advanced correlator
\begin{align}
\hat G_{ar} (\hat\omega) = 3\frac{\Phi^{\rm ig(3)}(-\hat\omega)} {\Phi^{\rm ig(0)}(-\hat\omega)} - i\hat\omega^3,
\end{align}
which obeys the Onsager relation $[\hat G_{ra}(\hat\omega)]^*= \hat G_{ar}(\hat\omega)$.

From \eqref{Grr_Gra}, via Legendre transformation (see \cite{Son:2009vu,Crossley:2015evo} for details), one can recover stochastic equation for heavy quark (here we set $2\pi \alpha^\prime =1$ for convenience):
\begin{align}
&\left[ \hat M \hat\omega^2 + \hat G_{ra}(\hat\omega) \right]\hat q_r^T(\hat\omega) = \hat \xi^T(\hat \omega), \qquad \qquad \langle \hat \xi^T(- \hat \omega) \hat \xi^T(\hat \omega) \rangle = \hat G_{rr}(\hat\omega), \nonumber \\
&\left[ \hat M \hat\omega^2 + \hat G_{ra}(\hat\omega) \right]\hat q_r^L(\hat\omega) = \hat \xi^L(\hat \omega), \qquad \qquad \langle \hat \xi^L(- \hat \omega) \hat \xi^L(\hat \omega) \rangle = \gamma^{-2}\hat G_{rr}(\hat\omega),
 \label{Langevin_eq_gen}
\end{align}
where $\hat \xi^T(\hat \omega) = -i \hat G_{rr}(\hat \omega) \hat q_a^T(\hat \omega)$, and $\hat \xi^L(\hat \omega) = -i \hat G_{rr}(\hat \omega) \hat q_a^L(\hat \omega)$. Here, \eqref{Langevin_eq_gen} generalizes the Langevin equation \eqref{Langevin_eq}, particularly the noise $\hat\xi^{T,L}$ becomes coloured. It is straightforward to rewrite these equations in terms of original variables (say, $\omega, q^{T,L}, \xi^{T,L}$) or in time domain, see \cite{Giecold:2009cg,CasalderreySolana:2009rm} for more details.

Below, we present our results for the retarded correlator $\hat G_{ra}(\hat \omega)$ either when $\hat \omega$ is small (analytically) or when $\hat \omega$ is arbitrary (numerically).

$\bullet$ {\bf Low frequency limit}

First, we consider the low frequency limit $\hat\omega/T \ll 1$ so that
\begin{align}
\Phi^{\rm ig}(\hat z, \hat\omega) = \Phi^{\rm ig}_0(\hat z, \hat\omega) + \lambda^1 \Phi^{\rm ig}_1(\hat z, \hat\omega) + \lambda^2 \Phi^{\rm ig}_2 (\hat z, \hat\omega) + \cdots,
\end{align}
where the bookkeeping parameter $\lambda \sim \hat\omega/T$. Besides regularity requirement at the horizon for $\Phi^{\rm ig}_n$ $(n=0,1,2,\cdots)$, we impose AdS boundary condition:
\begin{align}
\Phi^{\rm ig}_0(\hat z =0, \hat\omega) =1, \qquad \qquad \Phi^{\rm ig}_n(\hat z=0, \hat\omega)=0 \qquad {\rm for} ~~~ n\geq 1.
\end{align}
In the low frequency limit $\hat\omega/T<<1$, we are able to obtain the ingoing mode $\Phi^{\rm ig}$ up to third order $\mathcal O(\hat\omega^3)$:
\begin{align}
\Phi^{\rm ig}_0(\hat z, \hat\omega)&=1, \nonumber \\
\Phi^{\rm ig}_1(\hat z, \hat\omega)&=\int_0^{\hat z} \frac{x^2dx}{f(x)} \int_{z_h}^{x} \frac{2i\hat\omega}{y^3}dy=- \frac{i\hat\omega}{\pi T}\arctan(\pi T z) \nonumber \\
& \xrightarrow[]{\hat z \to 0} -i\hat\omega \hat z + \frac{1}{3}i\hat\omega \pi^2 T^2 \hat z^3 + \cdots, \nonumber \\
\Phi^{\rm ig}_2(\hat z, \hat\omega)&= \int_0^{\hat z} \frac{x^2dx}{f(x)} \int_{z_h}^{x} \left[-\frac{2}{y^2} i\hat\omega \partial_y \Phi^{\rm ig}_1(y,\hat\omega) + \frac{2}{y^3} i\hat\omega \Phi^{\rm ig}_1(y,\hat\omega) \right] dy\nonumber \\
&= - \frac{\hat\omega^2}{8\pi T^2} \left[-4\arctan(\pi T \hat z) + 4\left(\arctan(\pi T \hat z) \right)^2 + 2\log\frac{(1+\pi T\hat z^2)^2}{1+\pi^2 T^2 \hat z^2} \right] \nonumber \\
& \xrightarrow[]{\hat z \to 0} -\frac{1}{3}\hat\omega^2 \pi T \hat z^3+ \cdots, \nonumber \\
\Phi^{\rm ig}_3(\hat z, \hat\omega)& = \int_0^{\hat z} \frac{x^2dx}{f(x)} \int_{z_h}^{x} \left[-\frac{2}{y^2} i\hat\omega \partial_y \Phi^{\rm ig}_2(y,\hat\omega) + \frac{2}{y^3} i\hat\omega \Phi^{\rm ig}_2(y,\hat\omega) \right] dy \nonumber \\
& \xrightarrow[]{\hat z \to 0} \frac{1}{12}i\hat\omega^3 (-4+\pi -2\log2)\hat z^3 + \cdots.
\end{align}
Accordingly, the retarded two-point correlator $\hat G_{ra}$ is expanded as
\begin{align}
\hat G_{ra}(\hat\omega) = i\hat\omega \pi^2 T^2 -\hat\omega^2 \pi T + \frac{1}{4} i\hat\omega^3 (\pi - 2\log2)+ \cdots,
\end{align}
where the first two terms were reported before in \cite{Son:2009vu}, and the $\hat\omega^3$-term is consistent with relevant numerical result of \cite{Chakrabarty:2019aeu}.

$\bullet$ {\bf Beyond low frequency limit}

\begin{figure}[htbp!]
\centering
\includegraphics[width=0.48\textwidth]{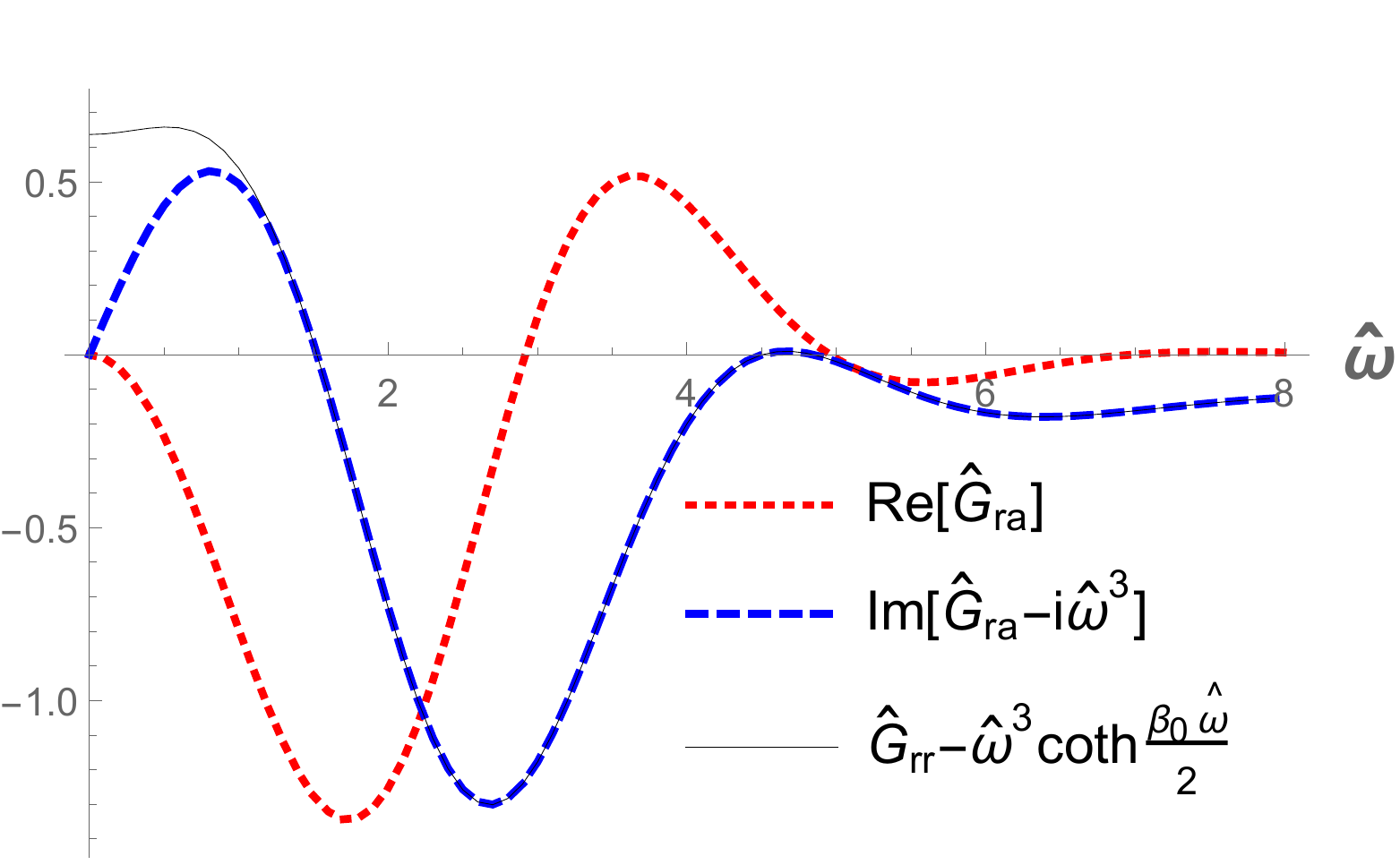}
\includegraphics[width=0.48\textwidth]{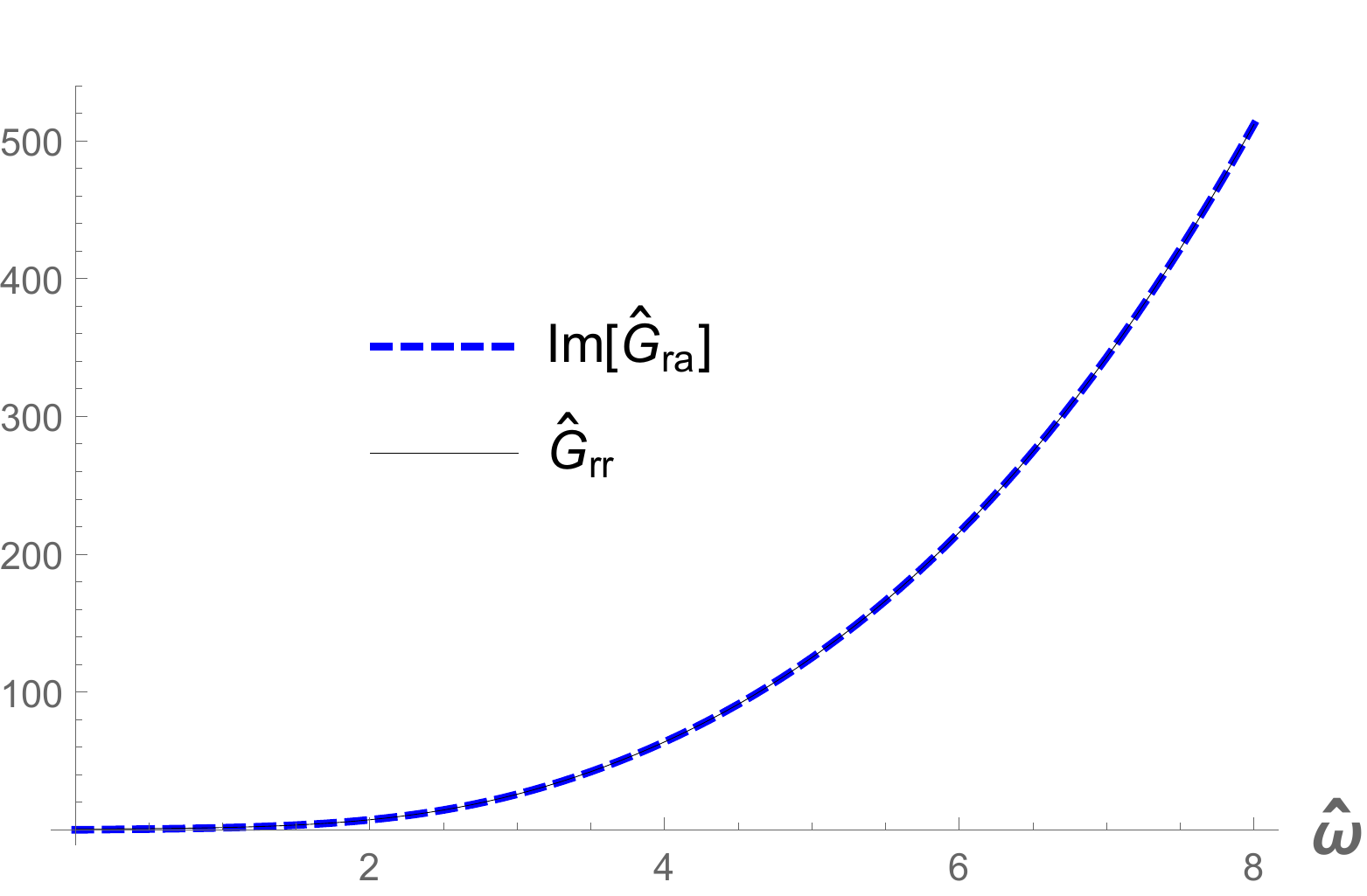}
\caption{Retarded and symmetric two-point correlators. Here, we set $\pi T =1$ for convenience.} \label{Gra-rr}
\end{figure}

For generic frequency $\hat\omega$, we numerically solve the EOM \eqref{eom_Phi} under regularity condition. In the ingoing EF coordinate, the analogue of \eqref{Phi_tilde_horizon} is
\begin{align}
\Phi^{\rm ig}(\hat z \to z_h,\hat\omega) = \Phi_h^0 + \Phi_h^1 (z_h -\hat z) + \Phi_h^2 (z_h -\hat z)^2 + \Phi_h^3 (z_h -\hat z)^3 + \cdots, \label{Phi_horizon}
\end{align}
where $\Phi_h^1, \Phi_h^2, \Phi_h^3, \cdots$ are fixed in terms of the horizon data $\Phi_h^0$. In practical calculation, we will take $\Phi_h^0=1$ so that the near horizon expansion \eqref{Phi_horizon} provides sufficient initial conditions for the EOM \eqref{eom_Phi}. Our numerical results for $\hat G_{ra}(\hat \omega)$ and $\hat G_{rr}(\hat \omega)$ are shown in Figure \ref{Gra-rr}, which are in perfect agreement with \cite{Gursoy:2010aa}. While ${\rm Re}[\hat G_{ra}]$ demonstrates a damped oscillating behavior as $\hat \omega$ gets increased, ${\rm Im}[\hat G_{ra}]$ (also $\hat G_{rr}$) shows a cubic growing behavior at large values of $\hat \omega$.

\subsection{Cubic action: non-Gaussianity} \label{cubic_action}

In this subsection, we compute cubic order effective action via replacing $\hat X_{T,L}$ in \eqref{S_NG_3rd} by the linearized string perturbation \eqref{X_TL_1} and then evaluating the radial integral. In the frequency domain, the cubic order action becomes
\begin{align}
S_{\rm NG}^{(3)} = - \frac{\beta \gamma}{2\pi \alpha^\prime} \int \frac{d\hat\omega_2 d\hat\omega_3}{(2\pi)^2} \delta(\hat\omega_1+\hat\omega_2+\hat\omega_3) \sum_{i=1}^{7} I_i(\hat\omega_1, \hat\omega_2, \hat\omega_3), \label{S_NG_3rd_omega}
\end{align}
where
\begin{align}
I_1& = \int_{0_1}^{0_2} d\hat z \frac{\gamma^2 \pi^2 T^2}{2} f(\hat z) \partial_{\hat z} \hat X_L^{(1)}(\hat z, \hat\omega_1) \partial_{\hat z} \hat X_L^{(1)}(\hat z, \hat\omega_2) \partial_{\hat z} \hat X_L^{(1)}(\hat z, \hat\omega_3), \nonumber \\
I_2& = \int_{0_1}^{0_2} d\hat z \frac{\pi^2 T^2}{2} f(\hat z) \partial_{\hat z} \hat X_L^{(1)}(\hat z, \hat\omega_1) \partial_{\hat z} \hat X_T^{(1)}(\hat z, \hat\omega_2) \partial_{\hat z} \hat X_T^{(1)}(\hat z, \hat\omega_3), \nonumber \\
I_3&= \int_{0_1}^{0_2} d\hat z \frac{-i\hat\omega_3\gamma^2}{2\hat z^2} (1-3\pi^2 T^2 \hat z^2) \partial_{\hat z} \hat X_L^{(1)}(\hat z, \hat\omega_1) \partial_{\hat z} \hat X_L^{(1)}(\hat z, \hat\omega_2)\hat X_L^{(1)}(\hat z, \hat\omega_3), \nonumber \\
I_4& = \int_{0_1}^{0_2} d\hat z \frac{i\hat\omega_3}{2\hat z^2} (1+\pi^2 T^2 \hat z^2) \partial_{\hat z} \hat X_T^{(1)}(\hat z,\hat\omega_1) \partial_{\hat z} \hat X_T^{(1)}(\hat z,\hat\omega_2) \hat X_L^{(1)}(\hat z, \hat\omega_3), \nonumber \\
I_5& = \int_{0_1}^{0_2} d\hat z \frac{-i\hat\omega_3 f(\hat z)}{\hat z^2 (1+\pi^2 T^2 \hat z^2)} \partial_{\hat z} \hat X_T^{(1)}(\hat z, \hat\omega_1) \partial_{\hat z} \hat X_L^{(1)}(\hat z, \hat\omega_2) \hat X_T^{(1)}(\hat z, \hat\omega_3), \nonumber \\
I_6& = \int_{0_1}^{0_2} d\hat z \frac{\hat\omega_2 \hat\omega_3 \gamma^2}{\hat z^2 (1+\pi^2 T^2 \hat z^2)} \partial_{\hat z} \hat X_L^{(1)}(\hat z, \hat\omega_1) \hat X_L^{(1)}(\hat z, \hat\omega_2) \hat X_L^{(1)}(\hat z, \hat\omega_3), \nonumber \\
I_7& = \int_{0_1}^{0_2} d\hat z \frac{\hat\omega_2 \hat\omega_3}{\hat z^2 (1+\pi^2 T^2 \hat z^2)} \partial_{\hat z} \hat X_L^{(1)}(\hat z, \hat\omega_1) \hat X_T^{(1)}(\hat z, \hat\omega_2) \hat X_T^{(1)}(\hat z, \hat\omega_3). \label{contour_integral_I1-7}
\end{align}
Our primary interest is to derive $S_{eff}^{(3)}$ up to first order in time derivative, which amounts to computing \eqref{contour_integral_I1-7} to the order $\mathcal{O}(\hat\omega^1)$. However, in order to discuss nonlinear KMS condition, we will also track part of $\mathcal{O}(\hat \omega^2)$-terms in $S_{eff}^{(3)}$, say $arr$-type terms. For a $\mathcal{PT}$-invariant theory in a thermal state, the KMS condition for generating functional sets four constraints among the full set of three-point correlators \cite{Wang:1998wg}: they involve all $\mathcal{O}(\hat\omega^1)$-terms in the cubic action, and additional $arr$-terms at the order $\mathcal{O}(\hat\omega^2)$, see appendix \ref{KMS_3point_review} for a summary on KMS constraints for three-point functions.

There is a subtle issue arising from the $\hat\omega$-expansion for $\hat X_{T,L}^{(1)}$ \eqref{X_TL_1}. While the function $A$ is regular over the contour, the function $B$ contains singularity (branch cut) near the horizon, due to the factor $e^{2i\hat\omega \chi(\hat z)}$. It is important to notice that near the horizon, the limit $\hat\omega \to 0$ and the limit $\epsilon \to 0$ of this factor $e^{2i\hat\omega \chi(\hat z)}$ do not commute:
\begin{align}
\lim_{\hat\omega \to 0} \lim_{\epsilon \to 0} e^{2i\hat\omega \chi(\hat z)} \neq
\lim_{\epsilon \to 0} \lim_{\hat\omega \to 0}  e^{2i\hat\omega \chi(\hat z)}, \qquad {\rm as} ~~ \hat z \to z_h + \epsilon. \label{non-commutativity}
\end{align}
If we expand the factor $e^{2i\hat\omega \chi(\hat z)}$ first in the limit $\hat\omega \to 0$, the nature of horizon singularity for this factor will be changed. Particularly, in order to guarantee $\hat \omega$-expansion to be convergent, the radius $\epsilon$ of the infinitesimal circle in Figure \ref{zcontour} should not be exponentially small \cite{Son:2009vu,CasalderreySolana:2009rm,Glorioso:2018mmw}:
\begin{align}
1 \gg \frac{\epsilon}{z_h} \gg e^{-l_{\rm mfp}T}, \label{epsilon_range}
\end{align}
where $l_{\rm mfp}$ is the typical time scale associated with the variations of the position $\hat q^{T,L}$. However, even when $l_{\rm mfp} T \to +\infty$ the condition \eqref{epsilon_range} seems to be in tension with the limit $\epsilon \to 0$. This subtlety has been avoided in the computation of quadratic order action $S_{\rm NG}^{(2), \rm P1}$, which is reduced into a boundary term without any expansion in small $\hat\omega$.

We have shown that the subtly \eqref{non-commutativity} is accidentally washed away in present work. In other words, in order to extract $\mathcal{O}(\hat\omega^1)$-terms and $\mathcal{O}(\hat\omega^2)$ $arr$-terms of \eqref{contour_integral_I1-7}, it becomes legal to expand the integrands of \eqref{contour_integral_I1-7} in small $\hat \omega$ before $\epsilon \to 0$ is taken. The demonstration is technical and is deferred to appendix \ref{non-commute}. We proceed by expanding integrands of \eqref{contour_integral_I1-7} in small $\hat \omega$ regime.

Up to $\mathcal{O}(\hat\omega^1)$, the functions $A, B$ \eqref{A_B} and their radial derivatives are expanded as:
\begin{align}
A(z,\hat\omega) = &1- \frac{i\hat\omega}{\pi T} \arctan(\pi T \hat z)+ \cdots, \nonumber \\
B(\hat z, \hat\omega) = &\left[ -\frac{1}{2} - \frac{2i}{\pi} \arctan(\pi T \hat z) -2i T \chi(\hat z) \right] + \hat\omega \left[- \frac{i}{2\pi T} \arctan(\pi T \hat z) - i\chi(\hat z) \right. \nonumber \\
&\left.  + \frac{2}{\pi} \arctan(\pi T \hat z) \chi(\hat z) + 2T (\chi(\hat z))^2 \right] + \cdots, \nonumber \\
\partial_{\hat z} A(\hat z, \hat\omega) = & -\frac{i\hat\omega}{1+\pi^2 T^2 \hat z^2}+ \cdots, \nonumber \\
\partial_{\hat z} B(\hat z, \hat\omega) = & \frac{2i \pi^2 T^3 \hat z^2}{1-\pi^4 T^4 \hat z^4} + \hat\omega \left[ \frac{i(1 + \pi^2T^2\hat z^2)}{2(1-\pi^4T^4 \hat z^4)} - \frac{2\arctan(\pi T \hat z)}{\pi (1-\pi^4 T^4 \hat z^4)} - \frac{2T (1+\pi^2 T^2 \hat z^2)}{1-\pi^4 T^4 \hat z^4} \chi(\hat z)\right] + \cdots,  \label{AB_hydro_scheme1}
\end{align}
where $\chi(\hat z)$ defined in \eqref{chi_on_contour} has explicit form:
\begin{align}
\chi(\hat z) = - \frac{\arctan(\pi T \hat z)}{2\pi T} - \frac{\log(1+\pi T \hat z)}{4\pi T} + \frac{\log(1-\pi T \hat z)}{4\pi T}, \qquad \hat z\in [0_1, 0_2]. \label{chi_expression}
\end{align}
Apparently, $\chi(\hat z)$ has a branch cut near the horizon and thus is a multi-valued function. In piecewise form, the function $\chi(\hat z)$ becomes
\begin{align}
\chi(\hat z) &= - \frac{\arctan(\pi T \hat z)}{2\pi T} - \frac{\log(1+\pi T \hat z)}{4\pi T} + \frac{\log(1-\pi T \hat z)}{4\pi T} - \frac{i}{2T}, \, \qquad \quad \hat z\in [0_1, z_h - \epsilon], \nonumber \\
\chi(\hat z) &= - \frac{\arctan(\pi T \hat z)}{2\pi T} - \frac{\log(1+\pi T \hat z)}{4\pi T} + \frac{\log(1-\pi T \hat z)}{4\pi T}\bigg|_{\hat z= z_h + \epsilon e^{i\theta}}, \qquad \theta\in [-\pi, \pi], \nonumber \\
\chi(\hat z) &= - \frac{\arctan(\pi T \hat z)}{2\pi T} - \frac{\log(1+\pi T \hat z)}{4\pi T} + \frac{\log(1-\pi T \hat z)}{4\pi T}, \qquad \qquad \qquad \hat z\in [0_2, z_h -\epsilon],  \label{chi_piecewise}
\end{align}

Since the integrands of $I_1, I_2$ only involve simple pole at $\hat z= z_h$ enclosed by the contour, they could be computed via residue theorem. The results are
\begin{align}
&I_1= \frac{3\gamma^2 \pi^2 T^3}{2} \left[ T \hat q_a^L(\hat\omega_1) \hat q_a^L(\hat\omega_2) \hat q_a^L(\hat\omega_3) + \hat\omega_1 \hat q_a^L(\hat\omega_2) \hat q_a^L(\hat\omega_3) \hat q_r^L(\hat\omega_1) \right], \nonumber \\
&I_2= \frac{3\pi^2 T^3}{2} \left[ T \hat q_a^L(\hat\omega_1) \hat q_a^T(\hat\omega_2) \hat q_a^T(\hat\omega_3) + \hat\omega_1 \hat q_a^L(\hat\omega_2) \hat q_a^T(\hat\omega_3) \hat q_r^T(\hat\omega_1) \right]. \label{I1_I2_result}
\end{align}

The integrals $I_{3,4,5}$ will involve logarithmic branch cut at $\hat z = z_h$, and will be computed by splitting the radial contour of Figure \ref{zcontour} as
\begin{align}
\int_{0_1}^{0_2} d\hat z = \int_{0_1}^{z_h -\epsilon} d\hat z + \int_{\mathcal{C}} d\theta + \int_{z_h-\epsilon}^{0_2} d\hat z, \label{contour_split}
\end{align}
where the integrals over the two horizontal legs will be partially cancelled. The integrals $I_{6,7}$ do not contribute at the order $\mathcal{O}(\hat\omega^1)$. For illustration we take the computation of $I_3$ as an example. For convenience, we split $I_3$ into three parts:
\begin{align}
I_3 =I_3^{(1)}\hat q_a^L(\hat\omega_1) \hat q_a^L(\hat\omega_2) \hat q_r^L(\hat\omega_3) + I_3^{(2)}\hat q_a^L(\hat\omega_1) \hat q_a^L(\hat\omega_2) \hat q_a^L(\hat\omega_3) + I_3^{(3)} \hat q_a^L(\hat\omega_1) \hat q_a^L(\hat\omega_2) \hat q_a^L(\hat\omega_3),
\end{align}
where $I_3^{(1)}$ and $I_3^{(2)}$ are computed directly via the residue theorem:
\begin{align}
I_3^{(1)}& = \int_{0_1}^{0_2}d\hat z \frac{2i\hat\omega_3 \gamma^2 \pi^4 T^6 \hat z^2 (1-3\pi^2 T^2 \hat z^2)}{(1-\pi^4 T^4 \hat z^4)^2} = \gamma^2 \pi^2 T^3 \hat\omega_3, \nonumber \\
I_3^{(2)}& = \int_{0_1}^{0_2}d\hat z \frac{\gamma^2 \pi^3 T^6 \hat\omega_3 \hat z^2 (-1 +3\pi^2T^2\hat z^2)}{(1-\pi^4T^4 \hat z^4)^2}\left[i\pi -2\arctan(\pi T \hat z) +\log(1+\pi T \hat z) \right] \nonumber \\
& = - \frac{1}{8} \gamma^2 \pi T^3 \hat\omega_3 \left[i +(4+2i)\pi -4i\log2 \right].
\end{align}
The third part is
\begin{align}
I_3^{(3)} = \int_{0_1}^{0_2}d\hat z \frac{\gamma^2 \pi^3 T^6 \hat\omega_3 \hat z^2 (1 -3\pi^2T^2\hat z^2)}{(1-\pi^4T^4 \hat z^4)^2} \log(1-\pi T\hat z),
\end{align}
where the integrand involves a branch cut at $\hat z = z_h$. Thus, $I_3^{(3)}$ will be computed by splitting the contour as in \eqref{contour_split}. The phase of the argument $\hat z$ is chosen as in \eqref{chi_piecewise}. The contribution from the horizontal legs is
\begin{align}
&\int_{0}^{1/(\pi T) -\epsilon} d\hat z \frac{\gamma^2 \pi^3 T^6 \hat\omega_3 \hat z^2 (1 -3\pi^2T^2\hat z^2)}{(1-\pi^4T^4 \hat z^4)^2} (-2i\pi) \nonumber \\
&= \frac{i\hat\omega_3 \gamma^2 T^2} {4\epsilon} + \frac{1}{2} i\hat\omega_3 \gamma^2 \pi T^3 \log \epsilon - \frac{1}{8}i\hat\omega_3 \gamma^2 \pi T^3 (-3+\pi +4\log2- 4\log(\pi T))+ \cdots,
\end{align}
where the infrared divergences will be exactly cancelled by the integral over the infinitesimal circle. In order to obtain the contribution from the infinitesimal circle, we turn to polar coordinate $\hat z = 1/(\pi T) + \epsilon e^{i \theta}$
\begin{align}
&\int_{\mathcal C} d\hat z \frac{\gamma^2 \pi^3 T^6 \hat\omega_3 \hat z^2 (1 -3\pi^2T^2\hat z^2)}{(1-\pi^4T^4 \hat z^4)^2} \log(1-\pi T\hat z) \nonumber \\
=& \int_{-\pi}^{\pi} d\theta\, i\epsilon e^{i\theta} \frac{\gamma^2 \pi^3 T^6 \hat\omega_3 \hat z^2 (1 -3\pi^2T^2\hat z^2)}{(1-\pi^4T^4 \hat z^4)^2} \left[ \log(\pi T \epsilon) +i\pi +i\theta \right]\bigg|_{\hat z \to 1/(\pi T) + \epsilon e^{i\theta}} \nonumber \\
= & - \frac{i\hat\omega_3 \gamma^2 T^2}{4\epsilon} - \frac{i\hat\omega_3 \gamma^2 \pi T^3}{2} \log \epsilon + \frac{1}{2} \gamma^2 \hat\omega_3 \pi^2 T^3 - \frac{1}{2} i\hat\omega_3 \gamma^2 \pi T^3 \log(\pi T) + \cdots.
\end{align}
So, the result for $I_3$ is
\begin{align}
I_3 =\gamma^2 \pi^2 T^3 \hat\omega_3 \hat q_a^L(\hat\omega_1) \hat q_a^L(\hat\omega_2) \hat q_r^L(\hat\omega_3) - \frac{1}{8} i\hat\omega_3 \gamma^2 \pi T^3 (3\pi -2) \hat q_a^L(\hat\omega_1) \hat q_a^L(\hat\omega_2) \hat q_a^L(\hat\omega_3). \label{I3_result}
\end{align}

The integrals $I_4, I_5$ could be computed in parallel with $I_3$. For brevity, we just record the final results:
\begin{align}
&I_4 =- \frac{1}{8}i\hat\omega_3 \pi T^3 (2+\pi) \hat q_a^L(\hat\omega_3) \hat q_a^T(\hat\omega_1) \hat q_a^T(\hat\omega_2), \nonumber \\
&I_5 = \pi^2 T^3 \hat\omega_3 \hat q_a^L(\hat\omega_2) \hat q_a^T(\hat\omega_1) \hat q_r^T(\hat\omega_3) - \frac{1}{4}i\hat\omega_3 \pi T^3(\pi -2) \hat q_a^L(\hat\omega_2) \hat q_a^T(\hat\omega_1) \hat q_a^T(\hat\omega_3). \label{I4_I5_result}
\end{align}

Finally, at the order $\mathcal{O}(\hat\omega^2)$ we only consider $arr$-terms in the cubic action. The integrals $I_1,I_2, I_5$ do not contribute. The contributions from $I_{3,4,6,7}$ are simply computed via residue theorem. The results are
\begin{align}
&I_3^{arr} = \gamma^2 \pi^2 T^2 \hat\omega_1 \hat\omega_3 \hat q_r^L(\hat\omega_1) \hat q_a^L(\hat\omega_2) \hat q_r^L(\hat\omega_3), \nonumber \\
&I_4^{arr} = \pi^2 T^2 \hat\omega_1 \hat\omega_3 \hat q_r^T(\hat\omega_1) \hat q_a^T(\hat\omega_2) \hat q_r^L(\hat\omega_3), \nonumber \\
&I_6^{arr} = \frac{\gamma^2 \pi^2 T^2}{2} \hat\omega_2 \hat\omega_3 \hat q_a^L(\hat\omega_1) \hat q_r^L(\hat\omega_2) \hat q_r^L(\hat\omega_3), \nonumber \\
&I_7^{arr} = \frac{\pi^2 T^2}{2} \hat\omega_2 \hat\omega_3 \hat q_a^L(\hat\omega_1) \hat q_r^T(\hat\omega_2) \hat q_r^T(\hat\omega_3). \label{arr_omega2}
\end{align}

%
%

Combining \eqref{I1_I2_result}, \eqref{I3_result}, \eqref{I4_I5_result} and \eqref{arr_omega2}, we obtain the cubic order SK effective action:
\begin{align}
S_{eff}^{(3)} = S_{eff}^{(3)LLL} + S_{eff}^{(3)LTT},
\end{align}
where
\begin{align}
S_{eff}^{(3)LLL} =& \frac{\beta \gamma}{2\pi \alpha^\prime} \int \frac{d\hat \omega_2 d\hat\omega_3}{(2\pi)^2} \left\{ \left[ -\frac{3}{2}\gamma^2 \pi^2 T^4 + \frac{1}{8} \gamma^2 \pi (3\pi -2) T^3 i \hat \omega_3 \right] \hat q_a^L(\hat \omega_1) \hat q_a^L(\hat \omega_2) \hat q_a^L(\hat \omega_3) \right. \nonumber \\
&\left. - \frac{5}{2} \gamma^2 \pi^2 T^3 \hat \omega_3 \hat q_a^L(\hat \omega_1) \hat q_a^L(\hat \omega_2) \hat q_r^L(\hat \omega_3)- \frac{3}{2} \gamma^2 \pi^2 T^2 \hat \omega_2 \hat \omega_3 \hat q_a^L(\hat \omega_1) \hat q_r^L(\hat \omega_2) \hat q_r^L(\hat \omega_3) \right\}, \nonumber \\
S_{eff}^{(3)LTT} =& \frac{\beta \gamma}{2\pi \alpha^\prime} \int \frac{d\hat \omega_2 d\hat\omega_3}{(2\pi)^2} \left\{ \left[ -\frac{3}{2} \pi^2 T^4 + \frac{1}{8} \pi (\pi +2) T^3 i\hat \omega_1 + \frac{1}{4} \pi (\pi -2) T^3 i \hat \omega_3  \right] \right. \nonumber \\
&\left. \times \hat q_a^L(\hat \omega_1) \hat q_a^T(\hat \omega_2) \hat q_a^T(\hat \omega_3) - \frac{5}{2} \pi^2 T^3 \hat \omega_3 \hat q_a^L(\hat \omega_1) \hat q_a^T(\hat \omega_2) \hat q_r^T(\hat \omega_3) - \pi^2 T^2 \hat \omega_1 \hat \omega_3 \right. \nonumber \\
&\left. \times \hat q_r^T(\hat \omega_1) \hat q_a^T(\hat \omega_2) \hat q_r^L(\hat \omega_3) - \frac{1}{2}\pi^2 T^2 \hat \omega_2 \hat \omega_3 \hat q_a^L(\hat \omega_1) \hat q_r^T(\hat \omega_2) \hat q_r^T(\hat \omega_3) \right\}, \label{Seff_3rd}
\end{align}
where $\hat \omega_1 + \hat \omega_2 + \hat \omega_3 =0$ is assumed. Recall that at the order $\mathcal{O}(\hat\omega^2)$ we only track $arr$-type terms. \eqref{Seff_3rd} is the main result of this work.

The cubic order action \eqref{Seff_3rd} represents not only nonlinear interactions among noises (the $aaa$-type terms) but also nonlinear interactions between dynamical variable $\hat q_r$ and noise (the $aar$-type terms), which are usually not covered in stochastic formulation, such as \eqref{Langevin_eq}. Moreover, the $aar$-type terms are analogous to multiplicative noise in classical stochastic models \cite{Kamenev2011}. The $arr$-type terms represents nonlinear $(\partial_t q)^2$ correction to the Langevin equation \eqref{Langevin_eq}. However, presence of $aaa$-type terms or $aar$-type terms makes it inconvenient to convert (via Legendre transformation) the effective action \eqref{Seff_2nd} and \eqref{Seff_3rd} into a stochastic equation for heavy quark position \cite{Crossley:2015evo}. In fact, many problems, although initially formulated as (generalized) Langevin equations with certain stochastic forces, are indeed studied by converting into an action framework, see \cite{Kamenev2011}. It is thus interesting to explore physical consequences of these terms by computing loop corrections to various observables, along the line of \cite{Chen-Lin:2018kfl,Jain:2020hcu,Sogabe:2021wqk,Jana:2021niz}, which goes beyond the scope of present work.

{\bf Consistency check}

$\bullet$ $Z_2$ reflection symmetry

Under the exchange of $\hat q_1^{L,T} \leftrightarrow \hat q_2^{L,T}$, the SK effective action shall satisfy \cite{Crossley:2015evo}:
\begin{align}
\left(S_{eff}[\hat q_1^T, \hat q_1^L; \hat q_2^T, \hat q_2^L]\right)^* = -S_{eff}[\hat q_2^T, \hat q_2^L;  \hat q_1^T, \hat q_1^L], \label{Z2}
\end{align}
which is actually unitary requirement, representing a self-consistent condition of the SK formalism. Under the constraint \eqref{Z2}, in the effective action (in time domain), the coefficient of any term with even numbers of $a$-type fields should be purely imaginary, while the coefficient of a term with odd numbers of $a$-type fields must be purely real. Obviously, our results for quadratic and cubic actions pass this basic requirement.

$\bullet$ KMS constraints for three-point functions

%
%
%

From the perspective of string worldsheet, we work around a thermal state (an emergent one on the worldsheet) with constant temperature, say \eqref{Schw_metric} or \eqref{EF_metric}. As analyzed in appendix \ref{S_NG_2nd_3rd_Schw}, while the quadratic bulk action \eqref{S_NG_2nd} preserves worldsheet $\mathcal{T}$-symmetry, the cubic bulk action \eqref{S_NG_3rd} does explicitly break it. On the other hand, the KMS constraints \eqref{KMS_3point} become applicable only if the theory is $\mathcal{T}$-invariant \cite{Crossley:2015evo}. Thus, it is reasonable to expect that the cubic order results \eqref{Seff_3rd} will not satisfy equilibrium KMS constraints \eqref{KMS_3point}, which we examine now.

Comparing \eqref{Seff_3rd} with \eqref{W_cub}-\eqref{HK_notation}, we can read off three-point correlators for medium force $\hat{\mathcal F}$. Let us take the $LLL$-sector as an example (see appendix \ref{KMS_3point_review} for nation convention):
\begin{align}
&\frac{1}{3!}\hat G^{LLL}= - \frac{3\gamma^2 \pi^2 T^4}{2} + \frac{1}{8}i\hat\omega_3 \gamma^2 \pi T^3 (3\pi -2), \nonumber \\
&\frac{i}{2}\hat H_1^{LLL}= - \frac{5\gamma^2 \pi^2 T^3}{2}\hat\omega_1, \quad \frac{i}{2}\hat H_2^{LLL}= - \frac{5\gamma^2 \pi^2 T^3}{2}\hat\omega_2, \quad \frac{i}{2}\hat H_3^{LLL}= - \frac{5\gamma^2 \pi^2 T^3}{2}\hat\omega_3, \nonumber \\
&\frac{1}{2}\hat K_1^{LLL}= -\frac{3\gamma^2 \pi^2 T^2}{2} \hat\omega_2 \hat\omega_3, \quad \frac{1}{2}\hat K_2^{LLL}= -\frac{3\gamma^2 \pi^2 T^2}{2} \hat\omega_1 \hat\omega_3, \quad \frac{1}{2}\hat K_3^{LLL}= -\frac{3\gamma^2 \pi^2 T^2}{2} \hat\omega_1 \hat\omega_2, \label{GHK_LLL}
\end{align}
where ``hatted'' $\hat G, \hat H, \hat K$ mean correlators of rescaled medium force $\hat{\mathcal F}$. In \eqref{GHK_LLL}, an overall factor $\beta\gamma/(2\pi \alpha^\prime)$ is ignored. Apparently, our results \eqref{GHK_LLL} do not satisfy \eqref{KMS_3point}. The same conclusion holds for $LTT$-sector.
 
%

\section{Summary and Outlook} \label{summary}

In this work, we considered nonlinear corrections to stochastic dynamics of a relativistic heavy quark in $\mathcal N=4$ SYM plasma, holographically described by a trailing string in AdS$_5$ black brane \cite{Herzog:2006gh,Giecold:2009cg,CasalderreySolana:2009rm}. Based on the holographic SK contour \cite{Glorioso:2018mmw}, we derived the SK effective action for the heavy quark, up to cubic order in quark's position.

At quadratic level, the SK effective action is computed first in the small $\hat \omega$ limit and then for arbitrary value of $\hat \omega$, perfectly satisfying ``worldsheet'' KMS condition, see \eqref{KMS_2point}. At cubic order, the SK effective action is computed perturbatively in small $\hat\omega$ regime. We observed that the KMS conditions for three-point functions of medium force are no longer obeyed due to steady state motion of the heavy quark. This becomes more transparent from string worldsheet's viewpoint: while the state of a trailing string results in an emergent static AdS black hole background on the string worldsheet, it also renders the cubic order bulk action to break worldsheet $\mathcal T$-invariance (see appendix \ref{S_NG_2nd_3rd_Schw}).

Another outcome of this work is on the technical side. We carried out a preliminary study over the subtle issue arising from non-commutative feature of the following two limits: the low frequency limit $\hat\omega/T \to 0$ versus $\epsilon \to 0$. Through brute-force calculations, we demonstrated that this subtlety is accidently absent in the calculation of SK effective action up to $\mathcal{O}(\hat\omega^1)$. However, this would not hold beyond first order in time derivative, and requires further investigation. We leave this open question for future study.

The present work focused on steady state motion of a Brownian particle in a static plasma medium. This can be extended in a few directions:

$\bullet$ First, it is interesting to consider a heavy quark moving in a viscous neutral flow background \cite{Abbasi:2012qz,Abbasi:2013mwa,Lekaveckas:2013lha,Reiten:2019fta}, which is dual to an open string probing slowly-varying AdS black brane \cite{Bhattacharyya:2008jc}. Such a study would be helpful in understanding fluctuation-dissipation relations for a non-equilibrium state \cite{CaronHuot:2011dr,Chesler:2013cqa,Hsiang:2020fbz}.

$\bullet$ Second, the present study would be extended to more realistic scenarios, say in the context of holographic models more closer to QCD plasma, as considered recently in \cite{Gursoy:2009kk,Gursoy:2010fj,Gursoy:2010aa,Cai:2012xh,Cai:2012eh,
Chernicoff:2012iq,Giataganas:2013zaa,Cheng:2014fza,Jahnke:2015obr,Chelabi:2015gpc,
Rougemont:2015wca,Li:2016bbh,Zhang:2018mqt,Chen:2019rez,Arefeva:2020bjk}.

$\bullet$ Last but not least, it is also interesting to consider spatially extended objects, and particularly when the low energy dynamics involves densities of conserved charges. One such example is the quantum critical system driven by a strong electric field \cite{Sonner:2012if}, aiming at revealing universal scaling behaviors for out-of-equilibrium situation. A second interesting instance is the steady state bubble nucleated within a chiral phase transition \cite{Bigazzi:2021fmq}. It is worth exploring nonlinear stochastic effects in both examples via holographic technique.


\appendix

\section{Perturbative $S_{\rm NG}$ in Schwarzschild coordinate} \label{S_NG_2nd_3rd_Schw}

In section \ref{string_perturbation}, the perturbative Nambu-Goto actions \eqref{S_NG_2nd} and \eqref{S_NG_3rd} are presented in the ingoing EF coordinate \eqref{EF_metric}, which makes the properties under $\mathcal T$-transformation obscure. In Schwarzschild coordinate \eqref{Schw_metric}, the perturbative Nambu-Goto actions \eqref{S_NG_2nd} and \eqref{S_NG_3rd} become:
\begin{align}
S_{\rm NG}^{(2)} = &-\frac{1}{2\pi \alpha^\prime} \int d\hat{t} \int_{0_1}^{0_2} \frac{d\hat{z}}{2\hat{z}^2}\left[-\frac{\gamma^2 } {f(\hat{z})} (\partial_{\hat{t}} \hat{X}_L)^2 + \gamma^2 f(\hat{z})(\partial_{ \hat{z}}\hat{X}_L )^2 \right. \nonumber \\
& \qquad \qquad \qquad \qquad \qquad \left.-\frac{1}{f(\hat{z})} ( \partial_{\hat{t}} \hat{X}_T )^2 + f(\hat{z}) (\partial_{\hat{z}}\hat{X}_T )^2\right], \label{S_NG_2nd_Schw}
\end{align}
\begin{align}
S_{\rm NG}^{(3)} = &-\frac{\gamma\beta}{2\pi \alpha^\prime}\int d\hat{t} \int_{0_1}^{0_2} \frac{d\hat{z}}{2\hat{z}^2f^2(\hat{z})}\bigg\{-\gamma^2 (\partial_{ \hat{t}} \hat{X}_L)^3-\gamma^2(\pi T \hat{z})^2f(\hat{z}) (\partial_{\hat{t}}\hat{X}_L)^2 \partial_{\hat{z}}\hat{X}_L \nonumber \\
&-\partial_{\hat{t}} \hat{X}_L(\partial_{\hat{t}}\hat{X}_T)^2-2(\pi T \hat{z})^2 f(\hat{z})\partial_{\hat{t}} \hat{X}_L\partial_{\hat{t}} \hat{X}_T \partial_{\hat{z}} \hat{X}_T +\gamma^2 f^2(\hat{z}) \partial_{\hat{t}} \hat{X}_L (\partial_{\hat{z}} \hat{X}_L)^2  \nonumber \\
&- f^2(\hat{z}) \partial_{\hat{t}} \hat{X}_L (\partial_{\hat{z}}\hat{X}_T)^2+(\pi T \hat{z})^2f(\hat{z})\partial_{\hat{z}}\hat{X}_L (\partial_{\hat{t}}\hat{X}_T)^2 +2f^2(\hat{z})\partial_{\hat{z}}\hat{X}_L\partial_{\hat{t}}\hat{X}_T\partial_{\hat{z}} \hat{X}_T \nonumber \\
&+\gamma^2(\pi T \hat{z})^2f^3(\hat{z})(\partial_{\hat{z}}\hat{X}_L)^3+(\pi T \hat{z})^2 f^3(\hat{z})\partial_{\hat{z}}\hat{X}_L(\partial_{\hat{z}}\hat{X}_T)^2 \bigg\}. \label{S_NG_3rd_Schw}
\end{align}
Now, from the worldsheet perspective, the $\mathcal T$-symmetry becomes more transparent. The second order action \eqref{S_NG_2nd_Schw} is invariant under the $\mathcal T$-transformation:
\begin{align}
\mathcal T: \qquad \hat t\to - \hat t, \qquad \hat X_{T, L}(\hat z, \hat t) \to \hat X_{T, L}(\hat z, -\hat t).
\end{align}
However, the cubic order action \eqref{S_NG_3rd_Schw} is not invariant under $\mathcal T$-transformation above. Moreover, it does not make sense to split $S_{\rm NG}^{(3)}$ into $\mathcal T$-invariant part and $\mathcal T$-breaking part, since it is the whole object $S_{\rm NG}^{(3)}$ that is a Lorentzian scalar. In some sense, the $\mathcal T$-breaking action $S_{\rm NG}^{(3)}$ is a reflection of steady state motion for the quark.

\section{Subtlety due to non-commutativity of $\epsilon \to 0$ versus $\hat \omega \to 0$} \label{non-commute}

In this appendix, we demonstrate that the non-commutativity issue of $\epsilon \to 0$ versus $\hat \omega \to 0$ becomes accidentally absent for present study. This validates the treatment of subsection \ref{cubic_action}.

Recall that each integrand in \eqref{contour_integral_I1-7} potentially has an oscillating factor $(1-\pi T \hat z)^{i\hat\omega/(2\pi T)}$, which results in the non-commutativity issue \eqref{non-commutativity}. With the expression \eqref{chi_expression}, the factor $e^{i\hat\omega \chi(\hat z)}$ can be rewritten as a product of an oscillating part and a regular part:
\begin{align}
e^{i\hat\omega \chi(\hat z)} = (1-\pi T \hat z)^{i\hat\omega/(2\pi T)} {\rm exp } \left\{ -\frac{i\hat\omega} {\pi T} \arctan(\pi T \hat z) - \frac{i\hat\omega}{2\pi T} \log(1+\pi T \hat z ) \right\},
\end{align}
where the singular portion $(1-\pi T \hat z)^{i\hat\omega/(2\pi T)}$ will not be expanded in small $\hat\omega$ limit before $\epsilon \to 0$. Once the integrals \eqref{contour_integral_I1-7} are done, we extract the low frequency limits.

We find it more convenient to rewrite $B$ of \eqref{A_B} as
\begin{align}
B(\hat z, \hat\omega) = B_1(\hat z, \hat\omega) + B_2(\hat z, \hat\omega) = B_1(\hat z, \hat\omega) + (1-\pi T \hat z)^{i\hat\omega/(2\pi T)} C(\hat z, \hat\omega),
\end{align}
where $B_1, C$ are regular functions over the domain enclosed by the contour
\begin{align}
&B_1(\hat z, \hat\omega)= \frac{1}{2}\coth\frac{\beta_0 \hat\omega}{2} \frac{\Phi^{\rm ig}(\hat z, \hat\omega)}{\Phi^{\rm ig(0)}(\hat\omega)}, \nonumber \\
&C(\hat z, \hat\omega) = - \frac{1}{1-e^{-\beta_0 \hat\omega}} \frac{\Phi^{\rm ig}(\hat z, -\hat\omega)}{\Phi^{\rm ig(0)}(-\hat\omega)} {\rm exp } \left\{ -\frac{i\hat\omega} {\pi T} \arctan(\pi T \hat z) - \frac{i\hat\omega}{2\pi T} \log(1+\pi T \hat z ) \right\}.
\end{align}
While the oscillating factor $(1-\pi T \hat z)^{i\hat\omega/(2\pi T)}$ is kept unexpanded, it is always legitimate to expand all regular parts, such as $A, B_1, C$, in small $\hat\omega$. In the small $\hat\omega$ limit, the functions $A, B_1, C$ and their radial derivatives scale as
\begin{align}
&A= \mathcal{O}(\hat\omega^0) + \cdots, \qquad \quad B_1= \mathcal{O}(\hat\omega^{-1}) + \cdots, \qquad \quad C= \mathcal{O}(\hat\omega^{-1}) + \cdots, \nonumber \\
&\partial_{\hat z} A = \mathcal{O}(\hat\omega^1) + \cdots, \qquad \partial_{\hat z} B_1= \mathcal{O}(\hat\omega^0) + \cdots, \qquad \partial_{\hat z} C= \mathcal{O}(\hat\omega^0) + \cdots. \label{ABC_omega_scaling}
\end{align}

Before calculating contour integrals \eqref{contour_integral_I1-7}, we make a formal analysis about each integrand. Recall that $A, B_1$ are regular, and $B_2$ has one branch cut at the horizon. So, in order to have non-vanishing contribution from each contour integral $I_i$ in \eqref{contour_integral_I1-7}, the integrand must contain one $B_2$-factor at least. Moreover, since $\partial_{\hat z} A \sim \mathcal{O}(\hat\omega^1)$, a term containing a factor $\partial_{\hat z} A \partial_{\hat z} A$ might be of order $\mathcal{O}(\hat\omega^2)$.
%

We take $I_3$ as an example to illustrate the computation. In terms of $A, B_1, B_2$, the integral $I_3$ becomes,
\begin{align}
I_3 = \int_{0_1}^{0_2} d\hat z & \frac{-i\hat\omega_3 \gamma^2}{2\hat z^2} (1-3\pi^2 T^2 \hat z^2) \left[2\partial_{\hat z}B_2(\hat z, \hat\omega_1) \partial_{\hat z} A(\hat z, \hat\omega_2)\, A(\hat z, \hat\omega_3) \hat q_a^L(\hat\omega_1) \hat q_r^L(\hat\omega_2) \hat q_r^L(\hat\omega_3) \right. \nonumber \\
& \qquad \qquad \left. + 2\partial_{\hat z}B_2(\hat z, \hat\omega_1) \partial_{\hat z} A(\hat z, \hat\omega_2) B_1(\hat z, \hat\omega_3) \hat q_a^L(\hat\omega_1) \hat q_r^L(\hat\omega_2) \hat q_a^L(\hat\omega_3) \right. \nonumber \\
& \qquad \qquad \left. + 2\partial_{\hat z}B_2(\hat z, \hat\omega_1) \partial_{\hat z} B_1(\hat z, \hat\omega_2) A(\hat z, \hat\omega_3) \hat q_a^L(\hat\omega_1) \hat q_a^L(\hat\omega_2) \hat q_r^L(\hat\omega_3) \right. \nonumber \\
& \qquad \qquad \left. + 2\partial_{\hat z}B_2(\hat z, \hat\omega_1) \partial_{\hat z} B_1(\hat z, \hat\omega_2) B_1(\hat z, \hat\omega_3) \hat q_a^L(\hat\omega_1) \hat q_a^L(\hat\omega_2) \hat q_a^L(\hat\omega_3) \right. \nonumber \\
& \qquad \qquad \left. + \partial_z A(\hat z, \hat\omega_1) \partial_z A(\hat z, \hat\omega_2) B_2(\hat z, \hat\omega_3) \hat q_r^L(\hat\omega_1) \hat q_r^L(\hat\omega_2) \hat q_a^L(\hat\omega_3) \right. \nonumber \\
& \qquad \qquad \left. + \partial_{\hat z}B_1(\hat z, \hat\omega_1) \partial_{\hat z} B_1(\hat z, \hat\omega_2) B_2(\hat z, \hat\omega_3) \hat q_a^L(\hat\omega_1) \hat q_a^L(\hat\omega_2) \hat q_a^L(\hat\omega_3) \right. \nonumber \\
& \qquad \qquad \left. + 2\partial_{\hat z}A(\hat z, \hat\omega_1) \partial_{\hat z} B_1(\hat z, \hat\omega_2) B_2(\hat z, \hat\omega_3) \hat q_r^L(\hat\omega_1) \hat q_a^L(\hat\omega_2) \hat q_a^L(\hat\omega_3) \right. \nonumber \\
& \qquad \qquad \left. + \partial_{\hat z}B_2(\hat z, \hat\omega_1) \partial_{\hat z} B_2(\hat z, \hat\omega_2) A(\hat z, \hat\omega_3) \hat q_a^L(\hat\omega_1) \hat q_a^L(\hat\omega_2) \hat q_r^L(\hat\omega_3) \right. \nonumber \\
& \qquad \qquad \left. + \partial_{\hat z}B_2(\hat z, \hat\omega_1) \partial_{\hat z} B_2(\hat z, \hat\omega_2) B_1(\hat z, \hat\omega_3) \hat q_a^L(\hat\omega_1) \hat q_a^L(\hat\omega_2) \hat q_a^L(\hat\omega_3) \right. \nonumber \\
& \qquad \qquad \left. + 2\partial_{\hat z}B_2(\hat z, \hat\omega_1) \partial_{\hat z}A(\hat z, \hat\omega_2) B_2(\hat z, \hat\omega_3) \hat q_a^L(\hat\omega_1) \hat q_r^L(\hat\omega_2) \hat q_a^L(\hat\omega_3) \right. \nonumber \\
& \qquad \qquad \left. + 2\partial_{\hat z}B_2(\hat z, \hat\omega_1) \partial_{\hat z} B_1(\hat z, \hat\omega_2) B_2(\hat z, \hat\omega_3) \hat q_a^L(\hat\omega_1) \hat q_a^L(\hat\omega_2) \hat q_a^L(\hat\omega_3) \right. \nonumber \\
& \qquad \qquad \left. + \partial_{\hat z}B_2(\hat z, \hat\omega_1) \partial_{\hat z} B_2(\hat z, \hat\omega_2) \partial_{\hat z} B_2(\hat z, \hat\omega_3) \hat q_a^L(\hat\omega_1) \hat q_a^L(\hat\omega_2) \hat q_a^L(\hat\omega_3) \right]. \label{I3_AB1B2}
\end{align}
Thanks to the delta function $\delta(\hat\omega_1+\hat\omega_2+\hat\omega_3)$, the last term of \eqref{I3_AB1B2} does not contain the singular factor $(1-\pi T \hat z)^{i\hat\omega/(2\pi T)}$, and will be computed via the residue theorem. The rest terms in \eqref{I3_AB1B2} contain an overall singular factor $(1-\pi T \hat z)^{i\hat\omega/(2\pi T)}$, and could be schematically written as
\begin{align}
\mathcal{I} = \int_{0_1}^{0_2} \frac{d \hat z}{\hat z^2} (1-\pi T \hat z)^{\pm i \hat\omega/(2\pi T)} \mathcal{H}(\hat z, \hat\omega), \label{I_H}
\end{align}
where $\mathcal{H}(\hat z, \hat\omega)$ contains simple pole and/or second-order pole near the horizon. Note that from \eqref{I3_AB1B2} to \eqref{I_H}, we ignored purely numerical factor as well as potential powers of frequency $\hat\omega^m(m=1~ {\rm or} ~ 2)$. Therefore, for the purpose of evaluating $I_3$ up to $\mathcal{O}(\hat\omega^1)$, it is sufficient to compute \eqref{I_H} up to $\mathcal{O}(\hat\omega^1)$. For a generic function $\mathcal{H}(\hat z, \hat\omega)$ it is challenging to obtain analytical results for \eqref{I_H}. Consider Laurent expansion of $\mathcal{H}(\hat z, \hat\omega)$ near the horizon
\begin{align}
\mathcal{H}(\hat z, \hat\omega) = \sum_{n=-2}^{\infty} \mathcal{H}_{n}(\hat\omega) (1- \pi T \hat z)^n, \label{H_Laurent}
\end{align}
which has a convergence radius larger than $1/(\pi T)$. Therefore, in the domain enclosed by the radial contour, it is legal to represent $\mathcal{H}(\hat z, \hat\omega)$ by its Laurent expansion \eqref{H_Laurent}. Here, in the small frequency limit $\mathcal{H}_{n}(\hat\omega)$ is a Taylor series of frequency. So, the original task of computing \eqref{I_H} now boils down to calculating
\begin{align}
\mathcal{I}_n = \int_{0_1}^{0_2} \frac{d\hat z}{\hat z^2} (1-\pi T \hat z)^{\pm i\hat\omega/(2\pi T)} (1- \pi T \hat z)^n, \qquad n=-2,-1,0,1,\cdots, \label{In_expression}
\end{align}
which could be worked out analytically for generic $\hat\omega$. Afterwards, we extract the small frequency behavior of $\mathcal{I}_n$
\begin{align}
& \mathcal{I}_n = \mp \hat\omega  \frac{ _2F_1(2,n+1;n+2;1-\Lambda)}{(n+1) T} + \mathcal{O} (\hat\omega^2), \qquad \quad \, n \geq 1, \nonumber \\
& \mathcal{I}_n = \mp \frac{\hat\omega }{ \Lambda T}+\mathcal{O}(\hat\omega^2) \qquad \qquad \qquad \qquad \qquad \qquad \qquad n= 0, \nonumber \\
&\mathcal{I}_n = 2 i\pi \mp \frac{ \hat\omega }{\Lambda T} + \mathcal{O} (\hat\omega^2), \qquad \qquad \qquad \qquad \qquad \quad ~~ n=-1, \nonumber \\
&\mathcal{I}_n = 4 i \pi +\mathcal{O}(\hat\omega), \qquad \qquad \qquad \qquad \qquad \qquad \qquad \quad n=-2, \label{In_results}
\end{align}
where $\Lambda$ represents a UV cutoff near the AdS boundary. It is straightforward to check that the results \eqref{In_results} could be correctly obtained by first expanding the integrand of \eqref{In_expression} in small $\hat\omega$ and then doing the radial integral. However, this latter approach cannot correctly cover higher order terms omitted in \eqref{In_results}, which correspond to higher order terms in the SK action.

Now, we turn to the $arr$-terms at the order $\mathcal{O}(\hat\omega^2)$, which correspond to the first line and fifth line in \eqref{I3_AB1B2}. With the scaling \eqref{ABC_omega_scaling}, it is obvious that only the first term of \eqref{I3_AB1B2} is relevant. Moreover, only the lowest order results in \eqref{In_results} are required.

The main lesson from analysis above is as follows. In order to obtain SK effective action to the order $\mathcal{O}(\hat \omega^1)$, it is legal although accidentally to first expand the integrands (including $(1-\pi T \hat z)^{\pm i\hat\omega/(2\pi T)}$) in \eqref{contour_integral_I1-7} in small $\hat\omega$, and then implement the radial integral. While generically this might not be true for higher order terms, it does work for extracting $arr$-terms at the order $\mathcal{O}(\hat\omega^2)$.

\section{KMS conditions for three-point functions} \label{KMS_3point_review}

Consider a quantum many-body system in a thermal state, which also preserves $\mathcal {PT}$-symmetry. The KMS condition, satisfied by the generating functional $W$, sets constraints among three-point functions \cite{Wang:1998wg}. These constraints were also re-derived and extended to situation with multiple variables in \cite{Crossley:2015evo}. In this appendix, we review these constraints for completeness. Here, we closely follow subsection II~D of \cite{Crossley:2015evo}.
 
At cubic level, the generating functional can be written as
\begin{align}
W_{\rm cub} = i \left[ \frac{1}{3!} G^{lmn} \phi_a^l \phi_a^m \phi_a^n + \frac{i}{2} H^{lmn} \phi_a^l \phi_a^m \phi_r^n + \frac{1}{2} K^{lmn} \phi_a^l \phi_r^m \phi_r^n \right], \label{W_cub}
\end{align}
where the indices $l$ etc labels different bosonic Hermitian operators $\mathcal O^l$ and associated sources $\phi^l$. Each term in \eqref{W_cub} should be understood as a convolution in Fourier space, e.g.,
\begin{align}
G^{lmn} \phi_a^l \phi_a^m \phi_a^n = \int \frac{d\omega_2 d \omega_3}{(2\pi)^2} G^{lmn}(\omega_1, \omega_2, \omega_3)\phi_a^l(\omega_1) \phi_a^m(\omega_2) \phi_a^n(\omega_3), \quad \omega_1+ \omega_2 + \omega_3=0,
\end{align}
where spatial momenta are suppressed. The three-point functions $G_{\alpha_1 \alpha_2 \alpha_3}^{lmn} $, with $\alpha_{1,2,3}= r ~{\rm or}~ a$, are third order functional derivatives of $W_{\rm cub}$ with respect to sources $\phi_{r,a}^l$. Then, we have
\begin{align}
G^{lmn} = - G_{rrr}^{lmn}, \qquad  H^{lmn}= G_{rra}^{lmn}, \qquad  K^{lmn}= G_{raa}^{lmn}. \label{3-point_function}
\end{align}
To present KMS constraints for three-point functions, it is convenient to introduce the following notations:
\begin{align}
&H_3^{lmn} \equiv G_{rra}^{lmn}, \qquad  H_2^{lmn} \equiv G_{rar}^{lmn}, \qquad  H_1^{lmn} \equiv G_{arr}^{lmn}, \nonumber \\
&K_1^{lmn} \equiv G_{raa}^{lmn}, \qquad  K_2^{lmn} \equiv G_{ara}^{lmn}, \qquad  K_3^{lmn} \equiv G_{aar}^{lmn}. \label{HK_notation}
\end{align}
Note that in \eqref{3-point_function} and \eqref{HK_notation} variables $\omega$'s are further suppressed.
For a $\mathcal{PT}$-invariant theory, KMS condition applied to three-point functions implies
\begin{align}
&H_1 = \frac{i}{2}(N_3+N_2) K_1^*- \frac{i}{2}(N_2 K_3 + N_3 K_2), \nonumber \\
&H_2 = \frac{i}{2}(N_3+N_1) K_2^*- \frac{i}{2}(N_1 K_3 + N_3 K_1), \nonumber \\
&H_3 = \frac{i}{2}(N_1+N_2) K_3^*- \frac{i}{2}(N_1 K_2 + N_2 K_1), \nonumber \\
&G= \frac{1}{4} \left[ (K_1^* + K_2^* + K_3^*) + 2 N_2 N_3 {\rm Re} K_1 + 2 N_1 N_3 {\rm Re} K_2 + 2 N_1 N_2 {\rm Re} K_3 \right], \label{KMS_3point}
\end{align}
where for brevity the indices $l,m,n$ in the functions \eqref{3-point_function} and \eqref{HK_notation} are suppressed. Moreover, we introduced
\begin{align}
N_a \equiv \coth\frac{\beta_0 \omega_a}{2}, \qquad a=1,2,3.
\end{align}
The first three equations in \eqref{KMS_3point} satisfy permutation symmetry among indices $(123)$.

In order to guarantee regularity in the limit $\omega \to 0$, the first three of \eqref{KMS_3point} will involve some terms in \eqref{W_cub} with at least one time-derivative, while the last of \eqref{KMS_3point} will involve some terms in \eqref{W_cub} with at least two time-derivatives. This explains why we need to additionally track $arr$-type terms in \eqref{Seff_3rd} at $\mathcal{O}(\hat \omega^2)$.

\section*{Acknowledgements}

This work was supported by the Natural Science Foundation of China (NSFC) under the grant No. 11705037.

\bibliographystyle{utphys}
\bibliography{reference}
\end{document}